\documentclass[12pt,reprint,onecolumn]{revtex4-1}

\usepackage{color,amsmath,amsfonts,amssymb,upref,mathrsfs,bm}
\usepackage[pdftex]{graphicx}
\usepackage[tight,nooneline, bf, BF]{subfigure}
\usepackage{slashed}
\usepackage{epsfig}
\usepackage[english]{babel}

\usepackage{color}

\def\({\left(}
\def\){\right)}
\def\[{\left[}
\def\]{\right]}
\def\be{\begin{equation}}
\def\ee{\end{equation}}
\def\a{\alpha}
\def\b{\beta}
\def\g{\gamma}

\def\s{\sigma}

\def\sh{\sqrt\h}
\def\t{\tau}
\def\vt{\vartheta}
\def\h{\hbar}
\def\U{{\bm u}}
\def\vk{\varkappa}
\def\hb{\hat\beta}
\def\vb{\tilde\beta}

\def\q{{\bf q}}
\def\g{{\bf g}}
\def\f{{\bf f}}

\def\ps{{\bm \psi}}
\def\ph{{\bm \phi}}
\def\vph{{\bm \varphi}}
\def\vvph{\hat{\bm \varphi}}
\def\cvph{\tilde{\bm \varphi}}

\def\brr{}

\def\co{k}
\def\n{\overline n}
\def\vtn#1#2{\vartheta_{#2}^{(#1)}}

\def\psn#1{{\ps^{(#1)}}}
\def\phn#1{{\ph^{(#1)}}}

\def\vvphn#1{{\vvph^{(#1)}}}
\def\cvphn#1{{\cvph^{(#1)}}}

\def\hbn#1{{\hb^{(#1)}}}
\def\vbn#1{{\vb^{(#1)}}}

\def\cn#1#2{{c_{#2}^{(#1)}}}

\def\I{{\bm{ \mathcal I}}}
\def\P{{\bm{ \mathcal P}}}
\def\G{{\bm \Gamma}}
\def\Ps{{\bm \Psi}}
\def\Ph{{\bm \Phi}}
\def\H{{\bm{ \mathcal H}}}

\def\K{{\bm{ \mathcal K}}}
\def\hK{\hat{\bm{ \mathcal K}}}
\def\Kc{{ \mathcal K}}
\def\B{\bm{ \mathcal B}}
\def\Bc{{ \mathcal B}}

\def\T{\bm{ \mathcal T}}
\def\Tc{{ \mathcal T}}

\def\Om{{\bm \Omega}}

\def\Phn#1#2{{\Ph_{#2}^{(#1)}}}

\def\Bn#1{\B^{(#1)}}
\def\Bcn#1{\Bc^{(#1)}}
\def\Kcn#1{\Kc^{(#1)}}
\def\Kn#1{\K^{(#1)}}
\def\Kn#1{\K^{(#1)}}
\def\Kcn#1{\Kc^{(#1)}}

\def\wt{\widetilde}

\def\Ref#1{(\ref{#1})}
\def\={ \mathop{=}}
\def\seq{ \mathop{\simeq}}
\def\arg{ \mathop{\rm arg}}
\def\Im{ \mathop{\rm Im}}
\def\Re#1{ \mathop{\rm Re}\!#1 \,}
\def\exp#1{ \mathop{\rm exp}\nolimits\hskip-.7mm\left\{#1\right\}}
\def\sgn{ \mathop{\rm sgn}}

\newtheorem{thm}{Theorem}

\begin{document}

\title{
Modes transformation for a Schroedinger type equation: avoided and unavoidable level crossings
}

\author{Ignat Fialkovsky}
\email{ifialk@gmail.com}
\affiliation{CMCC-Universidade Federal do ABC, Santo Andr\'e, S.P., Brazil}
\author{Maria Perel}
\thanks{Corresponding author, electronic mail: m.perel@spbu.ru}
\affiliation{St. Petersburg State University, 7/9 Universitetskaya nab., St. Petersburg 199034, Russia}

\date{\today}

\begin{abstract}

An asymptotic approach for a Schroedinger type equation with a non selfadjoint slowly varying Hamiltonian of a special type is developed. The Hamiltonian is assumed to be the result of a small perturbation of an operator with a twofold degeneracy (turning) point, which can be diagonalized at this point. The non-adiabatic transformation of modes is studied in the case where two  small parameters are dependent: the parameter characterizing an order of the perturbation is a square root of the adiabatic parameter. The perturbation of the Hamiltonian produces a close pair of simple degeneracy  points. Two regimes of mode transformation for the Schroedinder type equation are identified: avoided crossing of eigenvalues, corresponding to complex degeneracy points, and an explicit unavoidable crossing (with real degeneracy points).

Both cases are treated by a method of matched asymptotic expansions in the context of a unifying approach. An asymptotic expansion of the solution near a crossing point containing the parabolic cylinder functions is constructed, and the transition matrix connecting the coefficients of adiabatic modes to the left and to the right of  the degeneracy point is derived.

Results are illustrated by an example:  fermion scattering governed by the Dirac equation.
\end{abstract}

\maketitle
\newpage

\tableofcontents
\newpage


\section{Introduction}\label{sec:WKB}

The present paper is dedicated to investigation of the Schroedinger type equation:
\be
	\H\Ps= -i\h \frac{\partial \Ps(x)}{\partial x},
	\label{mainEq00}
\ee
where $\h$ is a small parameter and the Hamiltonian $\H$ may be non selfadjoint. Born and Fock \cite{born} showed in 1928 that under certain conditions on the Hamiltonian exact solution $\Ps$ can be approximated by a so called adiabatic mode $\Ps_{ad}$ with an error $O(\h)$. The latter is defined as
\be
\label{adiab}
\Ps_{ad} = \hat{\vph} \exp{\frac{i}{\h} \int\limits^x \hb dx' },
\ee
where $\hb$ is an eigenvalue  of $\H$ of multiplicity one,
 \be\label{Sch0}
 \H \hat{\vph} = \hb \hat{\vph},
 \ee
$\hat{\vph}$ is a corresponding eigenfunction with appropriate choice of the phase. This fact became known as the Quantum Adiabatic theorem. Numerous papers have been devoted to generalizations of this theorem, see reviews in \cite{teufel2003adiabatic} and \cite{hagedorn2007born}. The study of adiabatic approximation of solutions of (\ref{mainEq00}) was primarily stimulated by problems of molecular dynamics.

Special attention was paid to cases, where  adiabatic approximation fails because of local degeneration (or almost degeneration) of two eigenvalues
\be
  \hb_1(x_*) = \hb_2(x_*),
\ee
which induces non-adiabatic transitions between adiabatic modes taking place near $x_*$.  Two small independent parameters rule the problem: $\h$, which is an adiabatic parameter and the parameter characterizing the perturbation, which determines the discrepancy between almost degenerating eigenvalues.
The subject of study in such cases is the \emph{transition matrix} which connects adiabatic modes at different sides of the degeneracy point. It was first studied by Landau \cite{landau1965collected}, Zener \cite{zener_32}. The refined results were obtained by Stueckelberg \cite{stueckelberg_32} shortly afterwards. However, mathematically rigorous results came only much later. The difficulties arose, in particular, for such a relation between small parameters that enables exponentially small transitions between modes.  The account of exponentially small transitions between modes, which are given by expansions in powers of small parameter, is a complicated problem.   The first proof of the Landau-Zener formula in the simplest  case of dependent small parameters providing the entries of the transition matrix of order unity was given in \cite{hagedorn1991proof}. In  the case of arbitrary relation between small parameters the justification of formulas were given in \cite{joye1994proof}. We assume in the present paper that the small parameters are chosen dependent of each other to avoid difficulties of the exponentially small transitions.

The asymptotic results for Eq. \ref{mainEq00} with a selfadjoint Hamiltonian are termed time-adiabatic theory, or simply adiabatics, e.g. \cite{teufel2003adiabatic}. However, a small parameter may also exist in stationary Schroedinger equation and the relevant asymptotic methods are then named semiclassical or space-adiabatic methods. Generally speaking, the stationary equation is a multidimensional one and is not of the form (\ref{mainEq00}). However some semiclassical problems, which can be reduced to (\ref{mainEq00}) with the Hamiltonian satisfying conditions listed below, can be studied in the framework of our formalism.  Differential equations describing irregular waveguide problems  encountered in Mechanics and Electrodynamics  have one variable selected, it is the variable along the axis of a waveguide, it should be $x$. The differential equations are not normally represented in the form (\ref{mainEq00}) and contain the second derivatives in $x$, but can be reduced to the form (\ref{mainEq00}).  For slow enough inhomogeneity, they have  solutions in the form of \Ref{adiab},  named adiabatic.  As examples  of such problems, we can mention the problems of wave propagation in elastic or electromagnetic waveguides and  in the Timoshenko beam.   Adiabatic solutions fail if the phase velocities of two modes have a local point of the degeneration. In the field of wave propagation, the problems of the description of {\it interaction of modes} or {\it transformation of modes} or {\it coupling of modes} caused by the degeneracy point  are analogs of the problems of {\it non-adiabatic transitions} in Quantum Mechanics. Our aim here, in particular, is to generalize results elaborated in Quantum Mechanics to aforementioned problems, i.e. we suggest to study physical problems with a slow variation of parameters in one variable but without dissipation as a particular case of (\ref{mainEq00}) but with a non selfadjoint Hamiltonian operator.

To be specific, we assume that the non-selfadjoint Hamiltonian may be factorized
\be\label{factor}
  \H= \G^{-1} \hK,
\ee
with selfadjoint operators $\hK$ and $\G$ (all the conditions on $\H$ are scrutinized in Section \ref{sec:formul}).  Then Eq. \ref{mainEq00} is reduced to the form
\be
  \hK\Ps= -i\h \G\frac{\partial \Ps(x)}{\partial x}.
  \label{mainEq1}
\ee
In this case, for construction of adiabatic modes \Ref{adiab} we can use the same $\hb$, $\hat \vph$, which now can be considered as eigenvalues and eigenfunctions of a linear selfadjoint operator pencil
\be\label{pencil-oper}
  \hK \hat{\vph} = \hb \G\hat{\vph}.
\ee
Basing on \Ref{mainEq1} we can study  problems of mechanics or electrodynamics by analogy with time-adiabatic methods of Quantum mechanics.  The degeneracy points (often named the turning points) in the semiclassical theory and points of (avoided) crossing of energy levels are then all treated on the same footing.

At first time, the equation \Ref{mainEq1} was applied to an anisotropic electromagnetic waveguide in the monography \cite{felsen1994radiation}, where the operator $\K$ was a matrix differential operator with respect to variables transverse to $x$.
  Equation \Ref{mainEq1} has been already used by the some of us in several applied problems dealing with asymptotics near degeneracy points of different types: for Maxwell equations in the Earth-ionosphere waveguide \cite{perel1990overexcitation} (with a non-invertible $\G$), Maxwell equations in curvilinear coordinates near the smooth boundary of the convex body \cite{And-Zai-Per}, the Timoshenko beam equations \cite{perel2000resonance}, and elastic wave equations \cite{perel2005asymptotic}. It was also applied to investigation of liquid crystals \cite{Aksenova}.   The Dirac equation for quasiparticles in graphene is given in the form of \Ref{mainEq1}
  in Section \ref{app:dirac}.   Moreover, equation \Ref{mainEq1} may be used in the framework of $\mathcal{PT}$-symmetric quantum mechanics with $\G$ playing the role of $\mathcal P$-symmetry operator, see \cite{Bender2007} and reference therein.

One of the advantages of the formulation of all these problems in form \Ref{mainEq1} is that it allows for a clear and intuitive physical interpretation of processes of non-adiabatic transitions. Indeed, \Ref{mainEq1} possesses a conservation law
\be\label{conser}
  (\Ps,\G \Ps) = Const,
\ee
where, generally speaking, the quadratic form is not positively defined. For problems of wave propagation, it has a meaning of conservation of the time-averaged flux of energy, see the literature cited above. Then all modes can be divided into two groups depending on the sign of this constant, ones carrying energy to a degeneracy point, and others away from it. Such interpretation of the direction of mode propagation often does not coincide with that based on the sign of the phase velocity of modes.

\begin{figure*}
\centering
\subfigure[]
{   \includegraphics{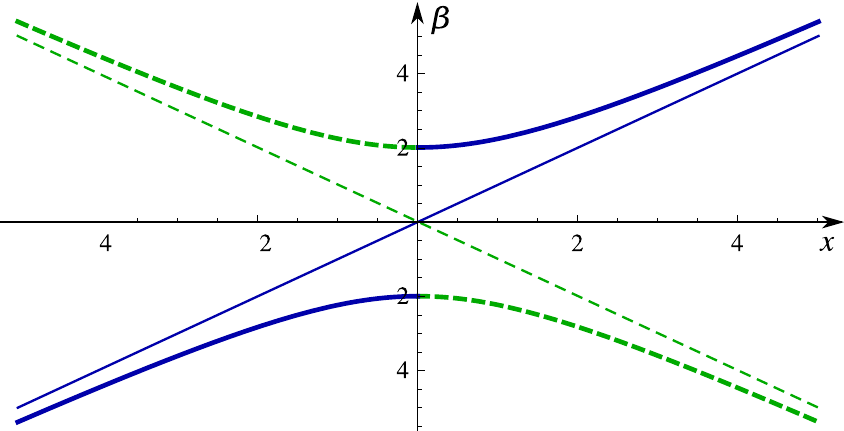}
\label{compl_TP}
}
\subfigure[]
{
\includegraphics{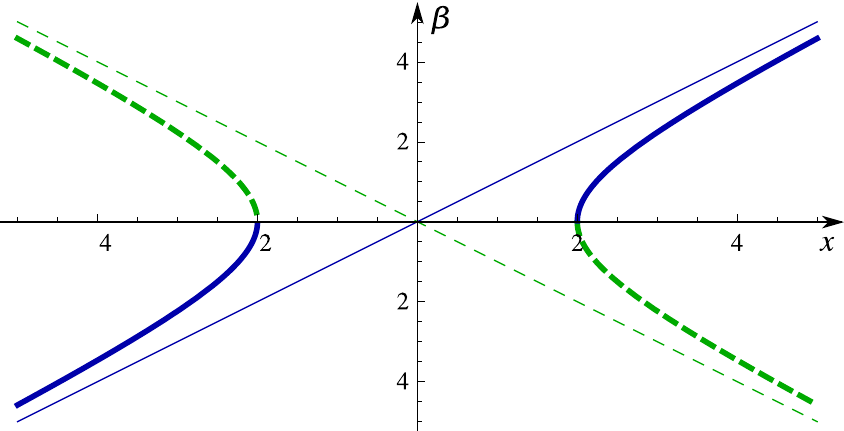}
\label{real_TP}
}
\caption{Eigenvalues vs. $x$ (in arbitrary units.) The non perturbed eigenvalues   are  shown by thin lines, while the exact ones $\hb$  by thick lines.   Dashed/solid lines (also blue/green on-line) distinguish modes number 1 and 2.
\subref{compl_TP} Avoided crossing: exact eigenvalues have two complex turning points.  \subref{real_TP} Unavoidable crossing: exact eigenvalues have two real turning points.
}
\end{figure*}

The problems of constructing of asymptotic solutions  as $\h \to 0$ in the presence of degeneracy points for the ODEs of second order and their generalizations to systems of ODEs has been studied in many papers;  see the books and  reviews \cite{berry72}, \cite{fedoryuk2012asymptotic}, \cite{olver2014asymptotics}, \cite{slavyanov1996asymptotic}, \cite{wasow2012linear} and references therein. All the methods applied to investigation of degeneracy points can be divided into three groups. First ones are the uniform methods, which work both in the vicinity of degeneracy points and away from them; see \cite{Buldyrev-Slavyanov}, \cite{Cherry}, \cite{langer1937connection}.  The second group comprises methods based on the Fourier representation of the unknown function, such as the Maslov method \cite{kucherenko1974asymptotics}, \cite{maslov2001semi} and the microlocal analysis  \cite{verdier_99}.  Methods based on local considerations in the vicinity of the degeneracy point with further matching of local solutions with adiabatic ones constitute the third group. There is the method of matched asymptotic expansions \cite{bender_book,wasow2012linear},  also called the boundary layer method  \cite{boundl_babich_79}. In the field of nonadiabatic transitions it was applied  in \cite{hagedorn1991proof}. This is the method we apply in the present paper.
The same method was used in our previous papers  \cite{perel1990overexcitation}, \cite{perel2000resonance}, \cite{perel2005asymptotic}, \cite{And-Zai-Per}
(although it was not always written explicitly), where different asymptotic situations for waveguide problems were considered.

Methods of distinguishing different types of asymptotic solutions also vary greatly. For simplest differential equation of the second order, the asymptotic  solutions near the degeneracy point   depend on the type of the $x$-dependence of eigenvalues. For a system of ordinary differential equations ($\hK$ is a matrix) and for more general cases, where $\hK$ is a differential operator,  some information about eigenfunctions is also necessary.  To build the classification of asymptotic cases, different approaches have been suggested in the literature.
 For equation in the form of \Ref{mainEq00} with a non selfadjoint Hamiltonian, the most typical cases were analyzed in \cite{buslaev_grigis_01,grinina2000solution} in assumption that a so called model equation is known.  However,  a recipe of reducing a concrete problem to a model equation was not suggested. The type of singularity of the  adiabatic coupling function should be prescribed in \cite{berry1993universal,betz2005precise} for classification of asymptotic cases.
A detailed classification of  points of level crossing  in terms accepted in molecular dynamics  is given in \cite{hagedorn1994molecular}. In our opinion, the reduction of a problem to \Ref{mainEq1} enables  an  easier classification of  asymptotic cases in terms of the matrix elements of $\hK$ and $\G$. One of the typical cases is considered in the present paper.

We assume that the operator $\hK(x)$ is  the result of a perturbation of an operator with a single degeneracy point, which is a point of crossing of two eigenvalues. It is well-known that if  $\G$ is an identity matrix, the crossing of eigenvalues turns under perturbation to an avoided crossing. It is a typical case for non-adiabatic transitions described by Landau-Zener formula. However  introducing  $\G$ we expanded the class of problems. The connection formulas should be revisited, and transition matrix should be generalized. A perturbation of operator with two eigenvalues crossing may cause now not only avoided crossing but unavoidable crossing too. Moreover the knowledge of eigenvalues behavior (for example, crossing or avoided crossing) is not sufficient for unique determination of the transition matrix. It is crucial whether there are two independent eigenfunctions at $x=0$ (in other words if the restriction of the operator $\G^{-1}\K$ to invariant subspace corresponding to $\b_1(x_*)=\b_2(x_*)$ is diagonalizable) or an eigenfunction and its associated. Particular cases of the treatment of the degeneracy points with linear independent eigenfunctions were presented in \cite{hagedorn1989adiabatic,perel2000resonance,And-Zai-Per}, and ones with linear dependent eigenfunctions in \cite{perel2005asymptotic}.
Here we consider the former case.

 We construct an inner and an outer formal asymptotic expansions,  show that the domains of their validity intersect, and match them to find the transition matrix.  The obtained expansions are asymptotic solutions, see subsection \ref{sec:wkb-valid}, \ref{sec:inner-ex}.  The outer expansion is the adiabatic one, which is well known for the selfadjoint Hamiltonians and was discussed  for the non selfadjoint ones in \cite{fedoryuk2012asymptotic} for linear systems of ordinary equations, and in \cite{nenciu1992adiabatic} for systems with large dissipation.  The equations for the leading term of the inner expansion can be interpreted as the model equation of \cite{buslaev_grigis_01,grinina2000solution}.
We show also that under the assumptions listed in Section \ref{sec:formul} these expansions are formal solutions of the equation \Ref{mainEq1}.

Our main aim is to suggest studying asymptotics of \Ref{mainEq1} for problems of mode transformation in waveguide problems, some of which were already studied by us. We give here one new example of the problem for the Dirac equation in the simplest case, where transverse to direction of propagation momentum is dependent on the adiabatic parameter. We intend to give example of elastic waveguides in the further paper.


\section{Statement of the problem and main result}\label{sec:stat-pr}
\subsection{Statement of the problem}\label{sec:formul}

We study asymptotic solutions of the following equation:
\be
	\hK(x,\h)\Ps(x,\h)= -i\h \G\frac{\partial \Ps(x,\h)}{\partial x},\qquad \hK(x,\h)\equiv \K(x)+\sqrt\h \B(x)
	\label{mainEq0}
\ee
in the Hilbert space.
We assume that $\K(x)$, $\B(x)$, $\G$ and its inverse, are operators that are selfadjoint for real $x$. The latter two of them are supposed to be bounded
, as well as $\K(x)-\K(0)$.
For simplicity, we consider $\G$ independent on $x$. (This condition may be removed but it is satisfied in all practical cases known to the authors.)

 We introduce two crucial assumptions concerning eigenvalues $\b$ and eigenfunctions $\vph$ of the spectral problem
\be
	 \K(x) \vph(x) = \b(x) \G \vph(x),
	\label{newEVP}
\ee
and an additional third assumption, which can be relaxed. {Note that eigenvalues of \Ref{newEVP} can be both real and complex, while eigenfunctions $\vph$ are $\G$-orthogonal, see Appendix \ref{app:eig-pr} for more details.}.

\begin{enumerate}
\item{
Two  eigenvalues of \Ref{newEVP},  $\b_1(x)$ and $\b_2(x)$, are degenerating at a single point $x=0$ and they stay real on the whole interval.
The  point $x=0$ is  a  degeneracy point of a crossing type, i.e.,
\be
	 \b_2(x) -  \b_1(x) \seq_{x\to0} 2 Q x,\qquad
	Q>0.
	\label{mb-cross}
\ee
Here $Q$ does not depend on $\h$, $Q \sim 1$ as $\h\to0$, and both  $\b_1$ and $\b_2$ are separated from the rest of the spectrum of  \Ref{newEVP} (if any) with a gap independent on $\h$.
}
\item{ The operator $\G^{-1}\K$ is diagonalizable in  the invariant subspace corresponding to $\b_1(0)=\b_2(0).$}

\item
    Operator families $\K(x)-\K(0)$ and $\B(x)$ are holomorphic in some domain including an interval  containing $x=0$.

We will use the following consequences of the third assumption (see, for example, \cite{kato2013perturbation}):

\begin{enumerate}
\item
Operators $\K(x)-\K(0)$ and $\B(x)$ can be expanded in the Taylor series with bounded operator coefficients.

\item
The eigenvalues $\beta_j(x)$ are holomorphic and can be expanded in the Taylor series.

\item
The eigenfunctions $\vph_j(x)$ are holomorphic and can be defined  at $x=0$ by continuity, the obtained eigenfunctions $\vph_1(0)$, $\vph_2(0)$ are $\G$-orthogonal:
\be
  (\vph_1(0),\G\vph_2(0))=0,
  \qquad \vph_j(0)\equiv \,\lim_{x\to 0} \vph_j(x),\quad j=1,2.
\ee

\end{enumerate}
\end{enumerate}

Our aim is to construct asymptotic outer (inner) expansions outside (inside) the neighborhood of the point $x=0$.  The outer expansion we also call {\it adiabatic} one or {\it adiabatic mode}.  After that we seek the transition matrix connecting two sets of adiabatic solutions at both sides of the point $x=0$.

We estimate the terms of obtained expansions and find conditions for them to have an asymptotic nature. The straightforward consequence of these estimates is the fact that obtained expansions are asymptotic solutions.
We call $\Ps_a^{(N)}(x,\h)$ {\it an asymptotic solution of order $N$} in powers of a small parameter $\h^{\gamma}$, $\gamma>0$ on an interval of $x$ if it satisfies \Ref{mainEq0} up to the terms of order ${\scriptstyle \mathcal{O}}(\h^{\gamma N})$, i.e.,
\be\label{result-theor}
\hK\Ps_a^{(N)} + i\h \G\frac{\partial \Ps_a^{(N)}}{\partial x} = {\scriptstyle \mathcal{O}} (\h^{\gamma N})
\ee
uniformly with respect to $x$. If a partial sum of an asymptotic series is an asymptotic solution of order $N$ for any $N$, then we call the series {\it an asymptotic solution}.


\subsection{Main result}

Our mains result is based on construction of the adiabatic solution  of \Ref{mainEq0}, which is performed in close analogy to the case of the self-adjoint Schroedinger equation \Ref{mainEq00}, see Section \ref{sec:outer}. Our construction differs from the standard one by two facts. First, $\hat\K$ is itself a function of the small parameter  $\sh$, i.e., $\hat \K\equiv \hat\K(x,\sh)$, see \Ref{mainEq0}. Eigenfunctions and eigenvalues of $\G^{-1}\hat\K$ denoted  $\hat\vph$ and $\hb$, respectively, also depend on $\sh$ and are found by the perturbation method from the eigenfunctions $\vph$ and eigenvalues $\b$ of $\G^{-1}\K$ in Sections \ref{sec:EVP-away}, \ref{sec:EVP-close}.  Therefore  we find asymptotic solution as expansion in terms of $\sh$ but not $\h$. Second feature is the presence of $\G$ in the right-hand side of \Ref{mainEq0}, which permits to express the eigenelements of the adjoint problem as $\vph^+ = \G \vph$, $\hat \vph^+ = \G \hat \vph$.

The leading order term (the asymptotic solution of order $N=0$) reads
\be
\Ps_{j\pm}^{(0)} = {\cal A}_{j\pm}
  \frac{ \vph_j(x)}{|{N}_j(x)|^{1/2}}
  \exp{i  \int\limits^x_{\Re \varkappa_{\pm}}\left( \frac{\hb_j (x')}{\h}   - \Im{  S_{jj}(x') } \right)dx' },\quad j=1,2,
 	\label{canon-res-0}
\ee
where ${\cal A}_{j\pm}$, $j=1,2$ are arbitrary constants. Adiabatic modes, generally speaking, differ on both sides of the  point $x=0$ and  the subscript $\pm$ indicates the side, where the mode is constructed. We note that the leading term \Ref{canon-res-0} depends on non-perturbed eigenfunctions $\vph_j$, see \Ref{newEVP}, and perturbed eigenvalues $\hb_j$. The $\G$ - normalization factor ${N}_j(x)$ and the diagonal conversion coefficient  $S_{jj}(x)$ are defined as
\be
	{N}_j(x)=( \vph_j(x),\G  \vph_j(x)),
\label{N}
\ee
 \be
  S_{kj}(x) = \frac{( \vph_k,\G d\vph_j/dx)}{N_k},
  \label{S11}
\ee
where  
the scalar product is that inherent from the Hilbert space, in which $\K$ acts.
Coefficients $S_{jj}(x)$ are connected with the Berry phase.

The choice of the lower limits of integration in \Ref{canon-res-0} influences the arbitrary constants. We take these limits equal to real parts of the degeneracy points for the perturbed operator $\G^{-1}\hat\K$,   which are as follows:
\be
\varkappa_{\pm}
	=
	\sh\(-b \pm \frac{|\Bc_{12}|}{\sqrt{N_1N_2}}\frac{e^{i\pi\(1+{\rm sgn}(N_1N_2)\)/4}}{d (\,\b_2-\b_1\,)/dx} \) \Big|_{x=0}, \quad  b = \frac{1}{d (\,\b_2-\b_1\,)/dx} \( \frac{\Bc_{22}}{  N_2} - \frac{\Bc_{11}}{  N_1}\)\Big|_{x=0},
  \label{limits-res}
\ee
where
\be
	{\cal B}_{ij}(x)\equiv \(\vph_i(x),{\bm{\mathcal B}}(x)\vph_j(x)\),\quad j,i=1,2.
	\label{B}
\ee
See details about the degeneracy points in Section \ref{sec:EVP-close}. Such choice of the lower limit is convenient although adiabatic solutions themselves are not applicable in the ${\cal O}(\sh)$ vicinity of $\Re \varkappa_{\pm}$,  as we will see later. However, calculating the integral of the eigenvalue $\hb_j$ over this region is possible, and to this end we find the expansion of eigenvalues $\hb_j$  in Appendix \ref{sec:EVP}. The sum of the first terms of expansions of eigenvalues contributing the leading order term of the adiabatic mode is denoted $\hb^{(main)}$ in the further text.




After these preliminary remarks and definitions we are able to formulate our main result, which concerns the matching of adiabatic modes.

\begin{thm}
Let $\hK=\K +\sh \B$, $\G$, and eigenvalues and eigenvectors of $\G^{-1}\K$ meet the conditions stated in Section \ref{sec:formul}. Then the equation \Ref{mainEq0} has an asymptotic solution $\Om$ of order 0 that has the following representations in terms of adiabatic modes to the left and to the right  of the turning point, $x=0$,
\begin{equation}
\begin{aligned}
\Om  =  k_1^- \Ps_{1-}^{(0)}+ k_2^- \Ps_{2-}^{(0)}, \,\quad   {\rm if} \quad x \lesssim -\h^{1/2-\gamma}, \qquad
\Om  =  k_1^+ \Ps_{1+}^{(0)}+ k_2^+ \Ps_{2+}^{(0)},  \,\quad  {\rm if} \quad x \gtrsim  \h^{1/2-\gamma},
\end{aligned}
\label{Om main}
\end{equation}
$\gamma>0$,
and the complex constants $k_j^{\pm}$, $j=1,2$ are connected by the transition matrix $\T$
\be
  \( \begin{array}{c}
      k_1^+\\
      k_2^+
     \end{array}
  \)
  =
  \T \( \begin{array}{c}
      k_1^-\\
      k_2^-
     \end{array}
  \),
\qquad
\T = \(\begin{array}{cc}
		 e^{ i\pi\nu} &
			\frac{i \sqrt{2\pi\nu} e^{ i\frac{\pi\nu}{2} +\nu - \nu \ln|\nu| } }{\Gamma(1-\nu)} \\
		\frac{ \sqrt{2\pi\nu}\, e^{ i\frac{\pi\nu}{2} -\nu + \nu \ln|\nu|}}{\Gamma(1+\nu)}&
			 e^{ i\pi\nu}
		\end{array}\).
\label{T-res}
\ee
\end{thm}
The dimensionless  parameter $\nu$ governing the result is as follows
\be
	\nu = i \frac{1}{2Q} \frac{\Bc_{12} \Bc_{21}}{ N_1 N_2} \Big |_{x=0}, \qquad
Q = d (\b_2-\b_1)/dx \Big |_{x=0}.
	\label{nu_gen-0}
\ee
If $N_1 N_2 =1$, \Ref{T-res} coincides with the Landau-Zener formula \cite{}. However for $N_1 N_2 =-1$, \Ref{T-res}  gives a matrix, which does not arise in considerations of self-adjoint Hamiltonians. These two cases describe two qualitatively distinct cases of eigenvalues asymptotics near $x=0$: avoided crossing (with complex turning points, Fig.~\ref{compl_TP}) and unavoidable crossing (with classical turning points, Fig.~\ref{real_TP}).
 Two distinct physical  processes correspond to these cases.

The transition matrix is  dependent on arbitrary constants  ${\cal A}_{j\pm}$. The form  \Ref{T-res} of $\T$  corresponds to the following choice of ${\cal A}_{j\pm}$:
\be\label{A-canon}
{\cal A}_{j\pm} = \exp{i(-1)^j \theta_a  + \frac{i}{2\h} \int\limits_{-\sh\, b}^{\Re \varkappa_{\pm}} \(\hb_1+\hb_2\) dx'},
\ee
\be
\theta_a =
    \arg \frac{\Bc_{12}}{N_1}+\frac\pi4\(1-{\rm sgn}(N_1N_2)\)
\Big|_{x=0}.
\label{theta-res}
\ee		
We call the adiabatic modes with such choice of ${\cal A}_{j\pm}$ the {\it canonical modes}.

It is important to note that although, generally speaking, $N_j$, $j=1,2$, may vanish in degeneracy points, the conditions 1)-2) from Section \ref{sec:formul} ensure that it does not happen. It is shown in Appendix \ref{app:eig-pr}.


\section{Example. Dirac equation in $2+1$ dimensions \label{app:dirac}}

The massless Dirac equation in $(2+1)$ - dimensional space with the external potential $U=U(x,y)$ for a stationary wave function $e^{iEt}\wt{\Ps}(x,y)$ reads
\be
	\(v_F {\bm \sigma}\cdot \hat{\bf p} +U\)\wt\Ps(x,y)=E\wt\Ps(x,y),
\label{dirac}
\ee
where
${\bm \sigma}=(\sigma_x,\sigma_y)$ is a pair of the Pauli $2\times 2$ matrices,
$\hat{\bf p}=-i\h \nabla$; for simplicity we put the Fermi velocity equal to one, $v_F=1$.

One of the physical problems described by \Ref{dirac} is the electron scattering on an electrostatic potential barrier inside graphene \cite{katsnelson2007graphene}. If the external potential depends on one variable only, $U=U(x)$, we can separate the $x$-derivative in the equation \Ref{dirac}, and use the Fourier transform for the $y$-dependence
\be\nonumber
{\Ps}(x,p_y)=\int\limits_{\mathbb{R}} d y e^{-ip_y y/\h}\wt{\Ps}(x,y).
\ee
Then, the case, where the electrons are incident almost perpendicularly to the potential barrier can be studied \cite{reijnders2013semiclassical,zalipPRB.91}. The interest in this case was invoked by the so-called Klein paradox \cite{klein1929reflexion} present in graphene, which is characterized by the unit probability of tunneling through the barrier, for applications in graphene, see \cite{katsnelson2006chiral}.

Close to normal scattering of electrons on a potential  is characterized  by small values of $p_y$.
In particular, if we assume that  $p_y$  is of order $\sh$, $p_y=\sh\, p$, $p=O(1)$, then we write  \Ref{dirac} in the form \Ref{mainEq0}
\be
	({\K}+\sh\B)\Ps=-i\h \G\frac{\partial\Ps}{\partial x}
\ee
with the Hermitian operators
\be
	      \K = \left(
           \begin{array}{cc}
             E-U(x) & 0 \\
             0 & E-U(x) \\
           \end{array}
         \right),\qquad
	      \B = \left(
           \begin{array}{cc}
             0& -i p \\
             ip &0\\
           \end{array}
         \right),\qquad
\G=      \left(
           \begin{array}{cc}
             0&1\\
             1&0 \\
           \end{array}
         \right).
\label{H0B gra}
\ee
 The operators $\K$, $\B$, and $\G$ act in the Hilbert space $l_2(\mathbb{C}^2).$
The conservation law \Ref{conser} in this case corresponds to the conservation of the $x$-component of the electron current $j^\mu$ \cite{Bogoliubov}, due to the  fact that $\G=\sigma_x$,
\be
	(\Ps,\G\Ps)  \equiv  j_x,
\ee
and can
 be both positive and negative.

The generalized eigenvalue problem $\K  \vph= \b\G \vph$ is solved by
\be
	 \b_j =(-1)^j (E-U(x)), \quad
 \vph_j= \frac1{\sqrt2}\left(
           \begin{array}{c}
             1 \\
             (-1)^j \\
           \end{array}
         \right),\qquad j=1,2,
\label{m vph gra}
\ee
and the $\G-$ normalization of modes reflects the electric charge current transferred by these modes, having opposite signs. So, in this case we have
\be
	N_1 N_2 = -1.
\ee

The degeneracy points $x=\vk$ are those where $\b(\vk)=0$, i.e.,
\be
	E= U(\varkappa).
\ee
Supposing that on the interval of interest there is only one such point $\vk=0$, we conclude that all the assumptions of the Section \ref{sec:stat-pr} are satisfied if the potential has a non vanishing first derivative at  $\vk$, $Q\equiv\frac{\partial U}{\partial x}(\vk)\ne 0$.

 For constructing the canonical modes, we calculate the matrix entries
\be
	\Bc_{12}=-\Bc_{21}=-ip, \qquad
	\Bc_{11}=\Bc_{22}=0,
\ee
and other necessary ingredients
\be
	S_{ij}=0,\qquad
	Q=U'(\varkappa),\qquad
	\nu=- i \frac{p^2 }{2 Q},
\ee
\be
	\theta_a=\arg \Bc_{12}+\pi/2=0, \qquad
	b=0,
\ee
\be
	\varkappa_\pm= \pm \sh \, p/Q.
\ee
We also note that in this case the eigenvalues are symmetric, $\b_1=-\b_2$.

The canonical modes are
\be
 {\Ps}_{j\pm}^{(0)} =  \vph_j (x)
  e^{i/\h \,\int\nolimits^x_{x^{}_{\pm}} \hb_j(x',\h) dx' },
 	\label{Psi-can-ell}
\ee
The eigenvalues of $\G^{-1}\hat\K$ are
\be
	\hb_{j}(x',\h)=(-1)^{j+1}  \sqrt{(E-U(x'))^2-p_y^2}, \quad j=1,2.
\ee
Outside the vicinity of the crossing point $x=0$, $E=U(0)$ we have the expansion:
\be
\hb_j(x',\h)= (E-U(x'))-\h\frac{  p^2}{ 2 (E-U(x')) } +\ldots.
\ee
If $x'$ is in the vicinity of the turning points of order ${\cal O}(\sh)$ inclusively, we get
\be
  \hb_1(x',\h) =
 - Q \,{\rm sgn}x \sqrt {(x')^2 -  \h p^2/Q^2 } + {\cal O}(\h)\label{hbet-pr-gra},
\ee
$\hb_2(x',\h)=-\hb_1(x',\h)$.
The transition matrix is given by \Ref{T} with $\nu$ from above.

Thus, our method can be applied readily to the description of the nearly normal scattering of electrons on an external potential. This case was investigated successfully in \cite{reijnders2013semiclassical,zalipPRB.91} by a different method, and it is a straightforward task to verify that our result for the transition matrix is in agreement with theirs.  We have an assumption that $\nu$ is of order one, but  result is valid for any $\nu$.


\section{Asymptotic derivation}

\subsection{Adiabatic (outer) expansion}\label{sec:outer}

To proceed with resolving the connection problem via the method of matched asymptotic expansions, in this section we construct adiabatic, or outer, expansions away from the degeneracy points of the original equation
\be
	\hK\Ps\equiv (\K + \sqrt\h \B)\Ps=-i\h\G\frac{\partial\Ps}{\partial x}.
	\label{mainEq}
\ee


\subsubsection{Construction of adiabatic expansion }\label{sec:wkb}
We search for an adiabatic expansion of a solution of \Ref{mainEq} in the  form of
\be\label{anzz}
	\Ps (x,\h) =\Ph(x,\h) \,e^{\tfrac{i}\h\int^x\vt(x',\h) \, dx'},
\ee
both $\Ph$ and $\vt$ are given by formal series in powers  of  $\sh$, respecting the order of magnitude of the perturbation
\begin{eqnarray}
	\Ph(x,\h)&=&\Phn0{}{}(x) +\sqrt\h\Phn{1}{}{}(x)  + \h\Phn2{}{}(x)  + \ldots,
		\label{anzatz01}\\
	\vt(x,\h)&=&\vtn0{} + \sqrt\h\vtn{1}{}(x)  + \h\vtn2{}(x)  +\ldots.
		\label{anzatz1}
\end{eqnarray}
The standard adiabatic approximation contains only an expansion of $\Ph$, while we also introduce the second expansion in the phase factor.  We are entitled to impose an additional condition
\be
	(\Phn0{}{},\G\Phn{n }{})=0, \qquad n\geq1.
\label{add_cond}
\ee
It fixes the arbitrariness of possible multiplication of the whole ansatz  \Ref{anzz} by an arbitrary series in $\h$.
This condition makes the representation (\ref{anzz})--(\ref{anzatz1}) unique. As we will see in the sequel, it also guarantees that the amplitude factor $\Ph$ depends  on the local properties of the medium only, while all the integral (nonlocal) ones are contained in the phase factor.

Following the perturbation method, we insert  (\ref{anzatz01}), (\ref{anzatz1}) into (\ref{mainEq}).  Equating the coefficients at equal powers of $\sqrt\h$, we obtain a  sequence of equations
\begin{eqnarray}
	(\K-\vtn0{}\G){\Ph}^{(0)}&=&
		0,\label{0-th}\\
	(\K-\vtn0{}\G){\Ph}^{(1)}&=&
		 -\B{\Ph}^{(0)}+\vtn{1}{}\G{\Ph}^{(0)},
		\label{1-th} \\
	(\K-\vtn0{}\G){\Ph}^{(2)}&=&
		 -\B{\Ph}^{(2)}+\vtn2{}\G{\Ph}^{(0)}+\vtn{1}{}\G{\Ph}^{(1)}-i\G\partial_x{\Ph}^{(0)},
		 \label{2-nd}\\
	\ldots&&\nonumber\\
	(\K-\vtn0{}\G){\Ph}^{(n)}&=&
		-\B{\Ph}^{(n-1)}+\vtn{n}{}\G{\Ph}^{(0)}
			+\sum_{i=1}^{n-1}\vtn{i}{}\G{\Ph}^{(n-i)}
		-i\G\partial_x {\Ph}^{(n-2)}.
		\label{n-th}
\end{eqnarray}

Aiming at constructing  the `first' mode, we choose  as the solution of the principal  order equation (\ref{0-th}) the first eigenfunction of $\K$ and corresponding eigenvalue
\begin{equation}
	\Phn{0}{1} =  \vph_1, \qquad \vtn0{1}= \b_1.
	\label{Phn0}
\end{equation}

We solve equations (\ref{1-th}), (\ref{2-nd}) and (\ref{n-th}) step by step.  All these equations
are solvable if their right-hand sides are orthogonal to the solution of the homogeneous equation (\ref{0-th}). This condition with account of (\ref{add_cond}) and \Ref{Phn0} yields
\begin{eqnarray}
&&\vtn1{1}=\frac{ \Bc_{\brr11}}{  N_1}, \quad
\vtn2{1}=\frac{( \vph_1, \B \Phn {1}{1}) +i ( \vph_1, \G \partial_x  \vph_1 )   }{   N_1}, \label{sec:t}\\
&&\vtn{n}{1}=\frac{( \vph_1, \B \Phn {n-1}{1}) +i ( \vph_1, \G \partial_x \Phn{n-2}{1} ) }{   N_1}, \quad  n>2. \label{b(n)}
\end{eqnarray}
We recall  here that the eigenvalues $ \b_j$, $j=1,2$,  are assumed real.

Taking into account (\ref{add_cond}), we write  the higher order approximations in the form
\be\label{first-ap}
	\Phn{n}{1}=\cn{n}{12} \vph_2 +  \vph_{1\perp}^{(n)},
	\qquad n\geq 1,
\ee
where $\cn{n}{12}$ is a scalar function of $x$, and  $ \vph_{1\perp}^{(n)}$ is $\G$-orthogonal to $ \vph_1$ and $ \vph_2$
\be\label{ort-def}
	( \vph_j, \G \vph_{1\perp}^{(n)})=0, \qquad j=1,2.
\ee
In the following sections we consider the degeneracy between $\b_1$ and $\b_2$, so we separate the  term proportional to $\vph_2$, because it contains the main singularity where  $\b_2$ is close to $\b_1$, see Section \ref{sec:wkb-valid}.
To find $\Phn{n}1$, we substitute (\ref{first-ap}) into (\ref{n-th}), calculate its scalar product with  $\vph_2$, take into account (\ref{ort-def}) and the orthogonality properties of eigenfunctions, see Appendix \ref{app:eig-pr}. We find
\begin{eqnarray}
&&\cn{1}{12}=\frac{ \Bc_{\brr21}}{( \b_1- \b_2)  N_2},
		\label{c-12-1}\\
&&\cn{n}{12}
	=\frac{( \vph_2, \B \Phn {n-1}{1}) - N_2 \sum_{i=1}^{n-1}\vtn{i}{1} \cn{n-i}{12}
		+ i ( \vph_2,\G \partial_x \Phn{n-2}{1} ) }{ ( \b_1- \b_2)  N_2}, n\ge 2.
					 \label{c-12-n}
\end{eqnarray}

To complete the construction of the adiabatic solution, we need to deduce the $\G$-orthogonal component $\vph_{1\perp}^{(n)}$, $n=1,2,\ldots$. However, as we shall see in the next section, it does not influence the transformation of the modes, at least in the principal order. Thus, we only need to show the possibility of its determination to prove that the recurrent system (\ref{0-th})-(\ref{n-th}) can indeed be solved  step by step. We rewrite equation (\ref{n-th})  as follows
\be\label{def-vperp}
	(\K-\b_1\G)\vph_{1\perp}^{(n)} =
		 \G \f^{(n)},
\ee
where $\G\f^{(n)}$ is the right-hand side of (\ref{n-th}).
 In view  of  the assumption of the presence of a finite gap between $\b_j,$ $j=1,2$ and the rest of the spectrum and the fact that $\f^{(n)}$ is $\G$-orthogonal to $\vph_j$, $j=1,2$, equation \Ref{def-vperp} has a single solution $\vph_{1\perp}^{(n)}$, see Appendix A, (property 5). Thus, all the terms of the formal series  can be constructed.

In the case of a purely discrete spectrum of $\K$, the first order approximation $\vph_{1\perp}^{(1)}$ has the  form of
\be
	\vph_{1\perp}^{(1)}= \sum\limits_{j\ne1,2} c_{1j}^{(1)} \vph_j = \sum\limits_{j\ne1,2} \frac{ \Bc_{\brr j1}}{( \b_1- \b_j)  N_j} \vph_j,
\ee
where the summation ranges  all  the modes except for the first two.

 In the $O(\sh)$ approximation, we have
\be
\Ps_1=  \( \vph_1	+ O(h^{1/2}) \) \exp{i \Theta_1 + \frac{i}\h\int^x_{x^*}
	    \(\hb_1^{(main)}(x,\h) -    S_{\brr11}\)dx
	     + O(h^{1/2})},
		\label{WKB1}
\ee
 where   $\Theta_1$ is  a constant phase factor, $\hb_1^{(main)}(x,\h)$ reads
\be\label{vb1}
\hb_1^{(main)} = \b_1+\sqrt\h \frac{ \Bc_{ 11}}{  N_1}
 +\h\[\frac{  \Bc_{ 21}  \Bc_{ 12}}{ ( \b_1- \b_2)   N_1  N_2 }
	+\frac{ (\vph_1,\B \vph_{1\perp}^{(1)})}{  N_1} \],
\ee
 it comprises the first terms of expansion of the eigenvalue $\hb_1$ in powers of $\sh$, which is valid in the area of applicability of adiabatic modes and which is found in Appendix \ref{sec:EVP} in \Ref{hb}. The coefficients $S_{ij}$ were defined in \Ref{S11}. The  imaginary part of $S_{jj}$  is the Berry phase \cite{Berry-phase}, while its real part fixes the normalization of the whole solution. Indeed, for smooth  $N_j=N_j(x)$  we find
\be
\Re \(S_{jj}\)
	= \frac{(\partial_x \vph_j, \G \vph_j)	
			+(\vph_j,\G \partial_x \vph_j) }
		{2(\vph_j,\G  \vph_j)}
	=  \frac{1}{2} \partial_x \ln {|(\vph_j, \G \vph_j)|},
	\label{Re S11}
\ee
and thus upon integration  and exponentiation in \Ref{WKB1}, at the upper limit of integration it gives exactly $|N_1(x)|^{-\tfrac12}$ and a constant at the lower one, so that the principal order of the adiabatic mode \Ref{anzz} is
 \be
\Ps_1^{(0)} =  e^{i\Theta_1}| N_1(x_*)|^{\tfrac12}\,
\frac{\vph_1(x)}{| N_1(x)|^{\tfrac12} }\exp{\frac{i}\h \int\limits^x_{x^*} \, \hb_1^{(main)}(x',\h) dx' - i \int\limits^x_{x^*}\Im{  S_{\brr11}(x') } dx' }.
 	\label{hWKB1-norm}
\ee
Thus, it can be made $\G$-normalized, $|(\Ps_1, \G \Ps_1)|=1+ O(\h)$, assuming the eigenfunctions are normalized in $x^*$. The overall sign of the normalization factor  $(\vph_1,\G \vph_1)$, however, cannot be fixed and represents intrinsic properties of solution. It is shown in Appendix \ref{app:eig-pr}(property 4) that $N_j(x), j=1,2$ is not equal to zero.

The structure of the amplitude $\Ph$ guarantees that under the transformation $\vph_j\to e^{i\sigma_j}\vph_j$, $j=1,2$,  it maps in the same way: $\Ph_j\to e^{i\sigma_j}\Ph_j$. And at the same time, the Berry phase $\int^x {\rm Im}   S_{\brr11} dx$ develops an opposite contribution under the same phase shift if $\s_j$ is not constant, $\s_j=\s_j(x)$,
\be
	\int^x_{x^*} {\rm Im} S_{\brr11} dx  \mathop\to_{\vph_j\to e^{i\sigma_j}\vph_j}
		\s_1(x) - \s_1(x^*) +\int^x_{x^*} {\rm Im}   S_{\brr11} dx.
\ee
So, the only ambiguity left in definition of $\Ps_j$ is an overall {\it constant} phase factor. It can be interpreted purely in terms of the lower limit of integration $x^*$, but  for simplicity of further analysis we introduced an additional parameter $\Theta_1$.

We  call the formal series constructed here {\it the adiabatic expansion} or  {\it adiabatics}. The principal term of the expansion  is named the adiabatic approximation or adiabatic mode.
The other solution, $\Ps_2$ is obtained by simply interchanging the indices $1\leftrightarrow 2$.


\subsubsection{The validity region of adiabatic solutions and the slow variable}
\label{sec:wkb-valid}

All the approximations $\Ps_j^{(n)}$, $n\ge 1$, are of order one if $\b_1 - \b_2$ is of order one. It follows from conditions in Section \ref{sec:formul}. Operators $\G$, $\B$ are bounded and norms of eigenfunctions $\vph_1$, $\vph_2$ and their derivatives are bounded. Norms of $\vph_{1\perp}^{(n)}$ are also bounded as it follows from Appendix \ref{app:eig-pr}.

According to \Ref{c-12-1}, \Ref{c-12-n} the higher order terms, $\Ps_j^{(n)}$, $n\ge 1$,  in expansion of $\Ps_j$, $j=1,2$ contain the singularity  near the degeneracy point $x=0$, where $\b_2(0)=\b_1(0)$. We study here how rapidly the degree of singularity increases with the growth of the order of approximation $n$.

The first important fact is that main singularity of $\Ps_j^{(n)}$ is contained in  the term $\cn{n}{12} \vph_2.$ The second necessary  fact is that $N_j$, $S_{jk}$, and $\Bc_{jk}$ are of order unity, as follows from our assumptions.
By induction on $n$ it is easy to show that for $n>1$ it holds
\begin{eqnarray}
&& \vtn{n}{1} \simeq \cn{n-1}{12} \,\frac{ \Bc_{12}}{  N_1}\, \simeq \,\frac{1}{x^{n-1}}, \qquad \qquad \\
&& \cn{n}{12} \simeq \frac{\cn{n-1}{12}}{ \b_1- \b_2} \, \frac{ \Bc_{22}}{  N_2}
		+ i \frac{\partial_x \cn{n-2}{12}}{  \b_1- \b_2} \, \simeq \,\frac{1}{x^{n}}.
\end{eqnarray}
Therefore, the higher terms of the expansions \Ref{anzatz01}, \Ref{anzatz1} can be estimated near the degeneracy point  as
\begin{equation}\label{est}
\h^{n/2} \vtn{n}{}(x) \,  =  \sqrt{\h}\,  O\(  \frac{\h^{(n-1)/2}}{x^{n-1}}\), \qquad \quad
\h^{n/2} \Phn{n} \, = \, O\(  \frac{\h^{n/2}}{x^{n}}\). \qquad \qquad \qquad
\end{equation}
They stay small and thus guarantee the asymptotic nature of expansions \Ref{WKB1} for
\be
	x\sim \h^{1/2-\gamma}\gg \h^{1/2}
\label{wkb_zone}
\ee
for any $\gamma>0$. This suggests to seek the \emph{resonant} or \emph{inner} expansion of (\ref{mainEq}) in the vicinity of $x=0$ in terms of the slow, or stretched,  variable
\be
	\tau=x/\sqrt\h.
	\label{slVar}
\ee

We note that the constructed asymptotic expansion is an asymptotic solution.
To obtain an asymptotic solution of $N$-th order, we need to take  partial sums of  $N$ terms of expansions \Ref{anzatz01}, \Ref{anzatz1} and insert these sums into \Ref{anzz}. Then we should substitute the result into   the equation \Ref{mainEq} and estimate the discrepancy. The terms of order up to $N$ are cancelled because of \Ref{0-th}-\Ref{n-th}, and the discrepancy comprises terms from the right-hand side of equations of order from $N+1$ till $2N$, which can be estimated as ${\cal{O}}(\h^{\gamma N})$, as it follows from the reasonings given above.

We can now analyze the structure of the outer asymptotic expansion in terms of $\tau$ for further constructing ansatz of inner expansion. Expanding  \Ref{b(n)} and \Ref{c-12-n} for small $x=\sh \t$ we deduce
\begin{eqnarray}
\label{bound-vtn}
&&\vtn{0}{}(\sh \t)  =  P^{(0)}_0 + \sh \t P^{(0)}_1  + (\sh \t)^2 P^{(0)}_2 + \ldots, \,\,\\
&&  \sh \vtn{1}{}(\sh \t)  =  \sh\(P^{(1)}_0 +\sh \t P^{(1)}_1 + (\sh \t)^2 P^{(1)}_2 + \ldots \),
\label{bound-vtn2}\\
&&   \h^{n/2} \vtn{n}{}(\sh \t) =  \sh\,{\t^{-n+1}} \(P^{(n)}_0 +\sh \t P^{(n)}_1 +(\sh \t)^2 P^{(n)}_2 + \ldots \),\quad n \ge 2, \\
&&\h^{n/2}\Phn{n}{}(\sh \t) =  \,{\t^{-n}} \(\U^{(n)}_0 + \sh \t \U^{(n)}_1 + (\sh \t)^2\U^{(n)}_2 + \ldots \), \quad n \ge 0.
 \label{bound-vtn3}
\end{eqnarray}
where $P^{(n)}_k$ and $\U^{(n)}_k$, $k=1,2$ are constant scalars and vectors correspondingly. After inserting \Ref{bound-vtn}-\Ref{bound-vtn3} into \Ref{anzatz01}, \Ref{anzatz1}, we can collect the terms with the same powers of $\sh$. Principal singularities of every term of the adiabatic expansion, i.e. the highest order terms in $(\sh \tau)^{-1}$, contribute to the first term of the inner expansion. The same is correct for the second terms and so on. We will find thus
\begin{eqnarray}
\frac{\vt(\sqrt{\h}\,\tau,\h)}{\sqrt{\h}} &=&
  \frac{P^{(0)}_0}{\sqrt{\h}}
  + \( \tau P^{(0)}_1+P^{(1)}_0+\tau\sum\limits_{n\ge2} \frac{ P^{(n)}_0}{\t^{n}} \)
    + \sqrt{\h}\( \t^ 2 P^{(0)}_2+\t P^{(1)}_1+ \t \sum\limits_{n\ge2} \frac{P^{(n)}_1}{\t^{n-1}} \)
\nonumber\\
&&    + \h\( \t^3 P^{(0)}_3+\t^2 P^{(1)}_2+\t^2\sum\limits_{n\ge2} \frac{P^{(n)}_2}{\t^{n-1}} \)
    + \ldots. \label{vt-re-exN}\\
\Ph(\sqrt\h\tau,\h) &=&
  \sum\limits_{n\ge0} \frac{ \U^{(n)}_0}{\t^{n}}
  +  \sqrt\h\,\t\sum\limits_{n\ge0} \frac{\U^{(n)}_1}{\t^{n}}
  +  \h\t^2\sum\limits_{n\ge0} \frac{\U^{(n)}_2}{\t^{n}}
  + \ldots \label{Ph-re-ex}
\end{eqnarray}
Upon integration over $\t$ (as required in \Ref{anzz}) the first term and the first parenthesis of \Ref{vt-re-exN} are non small for any values of $\t\gtrsim 1$, while the third and all consecutive are negligible if
\be
\t = \h^{-\gamma}, \quad 0 <\gamma < 1/6.
\ee
This defines the area of asymptotic nature of the expansion in term of $\t$, and thus also the applicability area of the inner solution (since such expansion is unique!)  Under this condition we can expand the exponent containing the third and higher terms of \Ref{vt-re-exN}, and obtain the following ansatz of the inner expansion:
\be
	\ps(\t,\h) = \ph(\t,\h) \,e^{\tfrac{i}\sh\int\limits^\t_{-b} \(a(\t) + \sh b(\t)\) d\t'},\quad  \ph(\t,\h)=\phn0(\t)+\sh\phn1(\t)+\h\phn2(\t)+\ldots,
	\label{anzatz2a}
\ee
The functions $a(\t)$, $b(\t)$ and $\phn{n}(\t)$, $n=1,2,\ldots$ will be found in Section \ref{sec:inner}.


\subsubsection{Canonical modes}\label{sec:canon-modes-I}

For the simplification of the further matching procedure and a simpler form of the transition matrix, we shall now introduce canonical modes by fixing arbitrary factors in the principal term \Ref{hWKB1-norm}. First of all, we note that in its phase
\be
\int\limits^x_{x^*} \hb_1^{(main)} dx
    \equiv \int\limits^x_{x^*} \(\frac{\hb_1^{(main)}+\hb_2^{(main)}}{2} + \frac{\hb_1^{(main)}-\hb_2^{(main)}}{2}\)dx
\ee
we can actually choose the lower limit of integration differently for each of terms, for it amounts only to a overall constant phase factor in $\Ps_{1}^{(0)}$.
As it will be evident from Section \ref{sec:can-matr}, the most simple form of the transition matrix is obtained for the {\it canonical modes} defined as
\be
\begin{aligned}
\Ps_{1\pm}^{(0)} =  \,e^{i\Theta_{1\pm}}
\frac{\vph_1(x)}{| N_1(x)|^{\tfrac12} }\exp{
  \frac{i}{2\h} \int\limits^x_{-\sh\, b} \(\hb_1^{(main)}+\hb_2^{(main)}\) dx'\right.\\
 \left. + \frac{i}{2\h}  \int\limits^x_{\Re \varkappa_{\pm}} \(\hb_1^{(main)}-\hb_2^{(main)}\) dx'
  - i \int\limits^x_{\Re\varkappa_{\pm}}\Im{  S_{11}(x') } dx' }.
 	\label{hWKB1-canon}
\end{aligned}
\ee
The subscript $\pm$ of $\Ps_{1\pm}^{(0)}$ and $\Theta_{1\pm}$ corresponds to the sign of $x$, i.e. to the side of the degeneracy point where the above asymptotics is used, and $\varkappa_{\pm}$ are degeneracy points, i.e., $\hb_1(\varkappa_{\pm}) = \hb_2(\varkappa_{\pm}),$ see details  in Section \ref{sec:EVP-close}.  Near these points the adiabatic expansion is not valid as we show in Section \ref{sec:wkb-valid}.
However,  eigenvalues are integrable, we will use below the expansion of $\hb_1$, which is found in Appendix \ref{sec:EVP-close}, of which we  need terms up to order ${\cal O}(\sh)$ inclusively.
The obtained formula \Ref{hWKB1-canon} may be written as \Ref{canon-res-0}, where $\hb^{(main)}_1$ approximates $\hb_1$ with an error of order ${\cal O}(\h)$.

 The second mode is obtained by interchanging the subscripts $1\leftrightarrow2$. The constant phase is given by
\be\nonumber
\Theta_{j\pm}=(-1)^j \theta_a,
\ee		
The choice of $\Theta_{j\pm}$ can be simplified if we also fix the phases of $\vph_j (0),$ $j=1,2$ in such a way that
\be\label{phase-condition}
	\frac{\Bc_{12}^{(0)}}{N_1^{(0)}}
	=\frac{\Bc_{21}^{(0)}}{N_2^{(0)}}.
\ee
Indeed, if we replace $\vph_1 (0)$ and $\vph_2 (0)$ by $e^{\frac{i\theta_a}2}\vph_1 (0)$ and $e^{\frac{-i\theta_a}2}\vph_2 (0)$, respectively, ${\Bc_{12}^{(0)}}$ is replaced by a real positive quantity for $N_1 N_2 >0$, and by a purely imaginary one for $N_1 N_2 <0$. In both cases, $\theta_a$ vanishes as follows from \Ref{theta-res}.
For such choice of eigenfunctions, we have simply
\be\label{Theta-can-short}
 \theta_a=0, \qquad \Theta_{j\pm} = 0,
\ee
which we assume in what follows.

Thus we complete the unambiguous definition of the adiabatic modes \Ref{hWKB1-canon}, which we named \emph{canonical modes}.

\subsubsection{Adiabatic modes rearrangement in the matching region}
\label{adia_rearr}

For the matching with the inner solution we shall now construct explicitly the principal term of the rearrangement of the adiabatic expansions on the boundary of their validity region:
\be
|\t| = \h^{-\gamma}, \qquad 0 < \gamma < 1/6.
\ee
For these values of $x=\sh \t$, the re-expansion of the phase of the first mode \Ref{hWKB1-canon} can be calculated  by using  \Ref{eig-root} and \Ref{hb-b-t-}, giving
\be\label{def-av}
\frac{\hb_1^{(main)}(\sh \t,\h)+\hb_2^{(main)}(\sh \t,\h)}{2} =
  \b_0 + \sh \hb_{av}^{(1)}(\t),
  \ee
  \be
\frac{\hb_1^{(main)}(\sh \t,\h)-\hb_2^{(main)}(\sh \t,\h)}{2} =
  \sgn(\t)\sh\sqrt {(\tau + b)^2 Q^2
		+{p^2}\,\sgn (N_1N_2)  }.
\ee
The formulas for $\b_0$ and $\hb_{av}^{(1)}(\t)$ see  in Section \ref{sec:EVP} in \Ref{b-0-def} and \Ref{b-av}.
Then changing the integration variable in the phase of \Ref{hWKB1-canon} and expanding the second integral at the upper limit by using \Ref{lar-t-as} we obtain
\be
\begin{aligned}
\frac{i}{2\h} \int\limits^x_{-\sh\, b} \(\hb_1^{(main)}+\hb_2^{(main)}\) dx'
  +\frac{i}{2\h}  \int\limits^x_{\Re \varkappa_{\pm}} \(\hb_1^{(main)}-\hb_2^{(main)}\) dx'
  - i \int\limits^x_{\Re\varkappa_{\pm}}\Im{  S_{jj}(x') } dx' \\
=  \frac{i}\sh\int\limits^{\t}_{-b}
	\(\b_0 + \sh \hb_{av}^{(1)}(\t')\) d\t'  - \frac{i Q (\t +b)^2}{2}  - \nu \ln|\sqrt{2Q} \tau|
 + i \zeta + O(\h^{1/2 - 3\gamma}),\label{rearr_ph}
\end{aligned}
\ee
where the remainder arose from integration of the higher terms, the term containing $S_{11}$ is of order $O(\h^{1/2-\gamma})$.   We introduce in \Ref{rearr_ph}  notation
\be
i \zeta =  - \frac{\nu}{2} + \frac{\nu \ln|\nu|}{2}.
\label{Theta1}
\ee
Taking the principal term of  the amplitude factor  in \Ref{hWKB1-canon}, we obtain
that on the different sides of the degeneracy point the first mode  is given by the same formula
\be
\Ps_{1\pm}^{(0)}(\sh\t,\h)
	=  \,e^{\tfrac{i}\sh\int\limits^\t_{ {-b}} \(\b_0 + \sh \hb_{av}^{(1)}\) d\t'}
		e^{i\zeta} \, |\sqrt{2Q}\t|^{-\nu} \, e^{-i \frac{ Q (\t+b)^2}{4}}
		\frac{\vph_1 (0)}{|N^{(0)}_1|^{1/2}}
	+  O(\h^{1/2 - 3\gamma}),
\label{Ps1p}
\ee
where the signs $+$ or $-$ correspond to positive or negative values of $\t$, respectively.
Interchanging the indices $1\leftrightarrow 2$ (which in \Ref{rearr_ph} accounts for changing the overall sign in all terms but the first one), for the second mode  we get
\be
\Ps_{2\pm}^{(0)}(\sh\t,\h)
	=   e^{\tfrac{i}\sh\int\limits^\t_{ {-b}}  ( \b_0 + \sh \hb_{av}^{(1)})  d\t'}
		e^{-i\zeta} \, |\sqrt{2Q}\t|^{\nu} e^{i\frac{Q (\t+b)^2}{4}}
		\frac{\vph_2 (0)}{|N^{(0)}_2|^{1/2}}
	+ O(\h^{1/2 - 3\gamma}).
\label{Ps2p}
\ee
We recall that $\nu$ is purely imaginary and the principal terms \Ref{Ps1p} and \Ref{Ps2p} are of order one in $\h$.


\subsection{Inner expansion}\label{sec:inner}

\subsubsection{Construction of asymptotic expansion}\label{sec:inner-ex}
We construct here a formal asymptotic expansion valid in the neighborhood of the degeneracy point and name it \emph{the inner} or \emph{ resonance expansion}.  To do this, we first express \Ref{mainEq} in terms of the slow variable $\tau$. Inserting expansions \Ref{expan-H-B} of $\K (\sh \tau)$ and $\B (\sh \tau)$  into  \Ref{mainEq}, we obtain
\be
	\(\Kn0+\sh (\tau \Kn1 +  \Bn0)
		+\h(\tau^2\Kn2 + \tau\Bn1)
			+ \ldots\)\ps=-i\sh\G\frac{\partial\ps}{\partial \t}.
	\label{resoEq}
\ee
Following ansatz \Ref{anzatz2a} its solution can be sought in the form of
\be
	\ps = \ph \,e^{\tfrac{i}\sh\int\limits^\t_{-b} \(\b_0 + \sh \hb_{av}^{(1)}\) d\t'},\quad  \ph=\phn0+\sh\phn1+\h\phn2+\ldots,
    \label{anzatz}
\ee
where $\b_0$ and $\hb_{av}^{(1)}$ are first terms of expansion of $(\b_1(\sh\t) + \b_2(\sh \t))/2$ in powers of $\sh$, see details in Appendix \ref{sec:EVP} in \Ref{b-0-def} and \Ref{b-av}, and the phase is chosen to simplify further matching with (\ref{Ps1p},\ref{Ps2p}).
Substituting \Ref{anzatz} into \Ref{resoEq} and equating the coefficients at equal powers of $\sh$, we obtain a sequence of equations
\begin{eqnarray}
(\Kn0-\b_0\G)\phn0&=&		0,\label{0-th-res}\\
(\Kn0-\b_0\G)\phn1&=&	(\hb_{av}^{(1)}\G - \tau \Kn1 - \Bn0)\phn0  -i\G\dot{\ph}^{(0)}, \label{st1-th-res} \\		
(\Kn0-\b_0\G)\phn2&=&   (\hb_{av}^{(1)}\G -  \tau\Kn1 - \Bn0)\phn1
 -(\tau^2\Kn2+ \tau\Bn1)\phn0 -i\G\dot{\ph}^{(1)}, \label{st2-nd-res}\\
	\ldots &&\nonumber
\end{eqnarray}
where the dot marks the derivative with respect to $\tau$, $\dot{f}\equiv \partial f/\partial  \tau$.
The recurrent system \Ref{0-th-res}--\Ref{st2-nd-res} is solved by analogy with the system obtained by the perturbation method for eigenvalues near degeneracy points, and their  solutions are given in terms of the same notation; see Subsection \ref{sec:EVP-close}.
 We note that the equations \Ref{0-th-tau}--\Ref{2-nd-tau} are transformed into \Ref{0-th-res}--\Ref{st2-nd-res} upon replacing $\hbn0$, ${\vph}^{(0)}$ $\vvphn1$  and $\hbn1$ by  $\b_0$, $\phn0$, $\phn1$, and $\hb_{av}^{(1)}-i \partial /\partial  \tau$, respectively. Thus we can repeat the  line of argument of Subsection \ref{sec:EVP-close}, also replacing $\a_{jk}^{(0)}(\tau)$ by
$a_{k}^{(0)}(\tau),$ $k=1,2$, to avoid a conflict of  notation.

Thus, the principal approximation of the amplitude, $\phn0$,  is given by the  formula
\be
    \phn0 = a_1^{(0)}(\tau)  \vph_1(0) + a_2^{(0)}(\tau) \vph_2(0),
	\label{phi0}
\ee
where the scalar coefficients $a_k^{(0)}(\tau), k=1,2$, satisfy a system of ordinary differential equations
\be
\begin{aligned}
	& 	(N_1^{(0)}\, (\,-i {\partial /\partial  \tau} + \hb_{av}^{(1)}\,)\,  - \tau \Kcn1_{11} - \Bcn0_{11})\, a_{1}^{(0)}
		+ (-\tau \Kcn1_{12} - \Bcn0_{12}) \, a_{2}^{(0)}=0,\\
	&   (-\tau \Kcn1_{21}- \Bcn0_{21}) \, a_{1}^{(0)}
	+(N_2^{(0)}\,(\,-i  {\partial /\partial  \tau} + \hb_{av}^{(1)}) - \tau \Kcn1_{22} - \Bcn0_{22}) \, a^{(0)}_{2}=0.
\end{aligned}
\ee
Using the definitions \Ref{not-Q}, \Ref{not-b} and  \Ref{b-av},  it can be written as
\be\label{zero-app-eq}
\begin{aligned}
	& 	-i \dot a_1^{(0)} +
		(\tau +b)Q   a_1 ^{(0)}		-   \frac{B_{12}^{(0)}}{N_1^{(0)}}   a_2^{(0)}  = 0,
	\\
	&	-i \dot a_2 ^{(0)}-
		\frac{B_{21}^{(0)}}{N_2^{(0)}} a_1 ^{(0)}		- (\tau +b) Q  a_2^{(0)} =0.
\end{aligned}
\ee
%
To solve this system, we first express $a_1^{(0)}$ from the second equation in the form of
\be
	a_1^{(0)}= -\frac{N_2^{(0)}}{B_{21}^{(0)}}\, \(i \dot a_2^{(0)} + (\tau +b) Q a_2^{(0)}\) 
	\label{a1}
\ee
and substitute it into the first one to obtain the following equation for $a_2^{(0)}$:
\be
	 \ddot a_2^{(0)}  + \((\tau +b)^2 Q^2 - i Q + \nu \s^2 \)a_2^{(0)}  =0,
	\label{a2}
\ee
here
\be
	\s =    e^{-i\pi/4}\sqrt{2Q}
	\label{sigma}
\ee
and $\nu$ is given by \Ref{nu_gen-0}.

We note that in terms of a new variable
\be
	t = \s(\tau +  b ), \label{def-t-sig}
\ee
the equation (\ref{a2}) reduces to the parabolic cylinder equation \cite{abramowitz}
\be
	\partial^2_t y(t)
	+ \(\frac12-\frac{t^2}4+\nu\) y(t)=0,
		\label{ParCyl}
\ee
where $a_2^{(0)}(\tau)= y(\s(\tau +  b ))$. The general solution to (\ref{ParCyl}) can be written as a linear combination
\be\label{a_2t}
	a_2^{(0)}(\tau) \equiv y(\s (\tau + b) )= A D_\nu(\s(\tau +  b ))+B D_\nu (-\s(\tau +  b )),
\ee
where $A$  and $B$ are arbitrary constants, and $D_\nu$ is the parabolic cylinder function in the notation of Whittaker \cite{Whittaker}. By using its property \cite{abramowitz}
$
	\partial_t D_\nu(t)+ (t/2) D_\nu(t)-\nu D_{\nu-1}(t)=0,
$
from (\ref{a1})  we derive the expression for the  first  coefficient in  \Ref{phi0}
\be\label{a_1t}
	a_1^{(0)}(\tau)= - i \frac{B_{12}^{(0)}}{ \s N_1^{(0)}}\,\, \( A D_{\nu-1}(\s(\tau +  b ))-B D_{\nu-1} (-\s(\tau +  b ))\),
\ee
where we took into account that
\be
\frac{\nu \s N_2^{(0)}}{\Bc_{21}^{(0)}} = \frac{\Bc_{12}^{(0)}}{ \s N_1^{(0)}}.
\ee

Using the formulas \Ref{anzatz}, \Ref{phi0}, \Ref{a_2t}, \Ref{a_1t} and \Ref{def-t-sig}, we obtain the principal approximation of the inner expansion
\begin{eqnarray}
	\ps(\tau,\h) & = & \phn0(\tau) \,e^{\tfrac{i}\sh\int\limits^\t_{-b} \(\b_0 + \sh \hb_{av}^{(1)}\) d\t'}  + O( \sh ),
\label{conn-ps} \\
\phn0(\tau) & = & -\frac{i \Bc_{12}^{(0)}}{\s N_1^{(0)}}
	\[A D_{\nu-1}(t)-B D_{\nu-1} (-t)\]  \vph_1(0) 
	+ \[ A D_\nu(t)+B D_\nu (-t)\]  \vph_2(0). \label{conn-ph0}
\end{eqnarray}

The process of constructing of the inner solution can be continued.
Representing higher approximation as $\phn{n}(\tau)= \sum_{j=1,2}a_j^{(n)} (\t) \vph_j(0)+{\bm \phi}^{(n)}_\perp(\tau)$, we  will obtain at each step the very same system \Ref{zero-app-eq} of two differential equations but now for  $a_j^{(n)}  (\tau)$, $j=1,2$, and with nontrivial right-hand side. The latter will be given by a particular linear combination of $a_j^{(l)}(\tau)$, $j=1,2,\perp$, $l \le n-1$ , found on the previous steps. After some algebraic transformations the free term can be written as a combination of $D_{\nu-1}(\pm t)$, $D_{\nu}(\pm t)$ with coefficients given by a polynomials in $\t$.  Then the solution for $a_j^{(n)} (\t)$, $j=1,2$ is also given by a similar combination. After some cumbersome but straightforward manipulations one can show that the relation
\be
	\h^{n/2} \phn{n} =  O(\h^{n/2} \tau^{3n})
\ee
holds for higher approximations, $|\t|\to\infty$; see Appendix \ref{app:est-app-inn}. Then the validity region of the inner solution is given by
\be
	|\tau| \sim \h^{-1/6 +\g'}
	\label{inters-zone0}
\ee
with arbitrary small $\g'>0$. Comparing with the validity region of the adiabatic solution \Ref{wkb_zone}, we observe that the intersection of these  regions or the \emph{matching region} is
\be
	|x| \sim \h^{1/2 - \g}, \quad 0 <\g< 1/6.
  \label{inters-zone}
\ee


 The straightforward consideration shows that the constructed inner expansion is an inner solution.
\subsubsection{Inner expansion in the matching region}\label{sec:inner-matching}
We derive now the principal term of the inner expansion, $\ps$, given by \Ref{conn-ps}, \Ref{conn-ph0} in the matching region, i.e., on the outskirts of the validity domain, as $\tau\equiv x/\sh \to \pm \infty$, but \Ref{inters-zone0} is still satisfied.

The asymptotics of  $D_\nu(t)$   is different in different sectors of the complex plane \cite{abramowitz}
\begin{eqnarray}
&& D_\nu(t) \seq_{|t|\to\infty} \quad  t^\nu e^{-\frac{t^2}{4}}(1+O(t^{-1})),\qquad\qquad\qquad \qquad \qquad \qquad
		 \arg(t)\in (-\frac{3\pi}{4}, \frac{3\pi}{4}),  \label{Dnu2}\\
&& D_\nu(t) \seq_{|t|\to\infty} \quad  t^\nu e^{-\frac{t^2}{4}}(1+O(t^{-1}))\quad
		+ \quad \xi_\nu e^{2 i\pi\nu} t^{-\nu-1} e^{ \frac{t^2}{4}}(1+O(t^{-1})),\nonumber\\
	&&	\qquad \qquad 	\qquad \qquad \qquad \qquad \qquad \qquad  \qquad \qquad \qquad	\qquad \qquad \qquad \arg(t)\in (\frac{\pi}{4}, \frac{5\pi}{4}),
\label{Dnu3}
\end{eqnarray}
where
$$
	\xi_\nu \equiv -\frac{\sqrt{2\pi}}{\Gamma(-\nu)} e^{-i\pi\nu},
$$
for the future use we note that $ \xi_{\nu-1}= \xi_\nu /\nu$.


To calculate the asymptotics of \Ref{a_1t}, \Ref{a_2t}, we need the following observation:
\be\label{D-1}
D_{\nu}(\pm\s (\t +b)) = e^{i\nu\,\arg{(\pm\s\t)}}\,   |\s\t |^{\nu} e^{-\frac{\s^2 (\t+b)^2 }{4}}
	+ O(\t^{-1})
\ee
is valid for $\arg{(\pm \s \t)} = -\pi/4$, or $\arg{(\pm\s \t)} = 3\pi/4$ and $|\t| \to \infty$. Here $b$ is a fixed   constant. We also took into account the fact that for imaginary $\nu$, the second term in \Ref{Dnu3} is of order $t^{-1}\sim \tau^{-1}$, while the exponent factor is a pure phase, $|\exp{\pm t^2/4}|=1$.

We need also the  asymptotics
\be\label{D-2}
D_{\nu-1}(\pm\s (\t +b)) = O(\t^{-1}),
\ee
valid for $\arg{(\pm \s \t)} = -\pi/4$, $|\t| \to \infty$ and
\be\label{D-3}
D_{\nu-1}(\pm\s (\t +b)) = e^{-i\nu\arg{(\pm\s\t)}}\, \xi_{\nu-1} e^{2 i\pi\nu} |\s\t |^{-\nu} e^{\frac{\s^2 (\t+b)^2 }{4}} + O(\t^{-1})
\ee
if $\arg{(\pm \s \t)} = 3\pi/4$, $|\t| \to \infty$.

 We proceed now with the asymptotics to the left of the turning point. We fix the arguments, so that for negative $\t$ and $|\t|>|b|$,
$$
\arg(\s(\tau+b)) = \arg (e^{-i{\pi}/{4}} \tau)= 3{\pi}/{4},\qquad
\arg (-\s(\tau+b))=-\pi/4.
$$
Taking into account \Ref{D-1}--\Ref{D-3}, for the amplitude of the principal term of the inner expansion \Ref{conn-ph0} as
${\t \to -\infty}$  we obtain
\begin{equation}\label{neg-as}
\begin{aligned}
\phn0 \=_{\t \to -\infty}
	-  \frac{ i \xi_{\nu-1} \Bc_{12}^{(0)}}{\s N_1^{(0)}}\,    A e^{i\frac{5\pi\nu}{4}}
 			\frac{ e^{\frac{\s^2(\t+b)^2}{4}}}{|\s\tau|^{\nu}}
			 \vph_1(0)
			 +
			 \(A e^{i\frac{3\pi\nu}{4}}   +  B e^{-i\frac{\pi\nu}{4}}\)
			 \frac {|\s\tau|^{\nu}}{ e^{\frac{\s^2(\tau+b)^2}{4}}}
				 \vph_2(0)  + O(\t^{-1}).
\end{aligned}
\end{equation}

Let us turn now to the asymptotics to the right of the turning point. For $\t>0$ we put
$$
\arg(\s(\tau+b)) = - \frac{\pi}{4},\qquad
\arg (-\s(\tau+b))=3\pi/4.
$$
Formulas \Ref{D-1}--\Ref{D-3} in this case yield an asymptotics as ${\t \to +\infty}$
\be
\begin{aligned}
\phn0 \=_{\t \to +\infty}
	 \frac{  i \xi_{\nu-1}\Bc_{12}^{(0)}}{ \s N_1^{(0)}}  \,  B e^{i\frac{5\pi\nu}{4}}
			 \frac{ e^{\frac{\s^2(\t+b)^2}{4}}}{|\s\tau|^{\nu}}
			 \vph_1(0)
 +  \(A e^{-i\frac{\pi\nu}{4}}   +  B e^{i\frac{3\pi\nu}{4}}\)
			\frac {|\s\tau|^{\nu}}{ e^{\frac{\s^2(\tau+b)^2}{4}}}
				 \vph_2(0)  + O(\t^{-1}).
\label{pos-as}				
\end{aligned}
\ee
In both cases,  \Ref{conn-ps} should be applied for the inner expansion.


\subsection{Transition matrix }\label{sec:T}
\subsubsection{Matching of inner and outer expansions}\label{sec:numbering}

The adiabatic solutions $\Ps_j$, $j=1,2$ can only be defined at one side of a degeneracy point at a time, and by passing over it the transformation of modes generally occurs. Let the asymptotics as $\h \to 0$ of an exact solution $\Om$ be a linear combination
\be
	\Om\seq_{x \ll -\sh}  k_1^- \Ps_1 + k_2^- \Ps_2,	
\ee
on one side of the degeneracy point, say, to its left, $x< 0$.
Then the aim of the {\it connection problem} is to derive the coefficients of the linear combination on the other side, for $x> 0 $,
\be
	\Om\seq_{x \gg \sh} k_1^+ \Ps_1 + k_2^+ \Ps_2.
\ee
as functions of $k_1^-$ and $k_2^-$. It is solved by the transition matrix $\T$ defined by the equality
\be
\(\begin{array}{c}
		 k_1^+\\
		 k_2^+
		\end{array}\)
= \T \(\begin{array}{c}
		 k_1^-\\
		 k_2^-
		\end{array}\).
\label{trMa}
\ee
To obtain it, we shall match the inner and outer expansions in the intersection of their validity zones, \Ref{inters-zone}.

Comparing \Ref{neg-as} and \Ref{pos-as} with \Ref{Ps1p} and  \Ref{Ps2p} correspondingly, we immediately recognize that
\begin{eqnarray}
 && {|N_2^{(0)}|^{-1/2}}	\psn0
	 \mathop{\simeq}_{\tau\to-\infty}
		 \co_1^- \Ps^{(0)}_{1- } +    \co_2^-\Ps^{(0)}_{2- },
	\label{ps-infty-}\\
&&	{|N_2^{(0)}|^{-1/2}} \psn0
	 \mathop{\simeq}_{\tau\to+\infty}
		 \co_1^+ \Ps^{(0)}_{1+ } +    \co_2^+\Ps^{(0)}_{2+ }, \label{ps-infty+}
\end{eqnarray}
where
\begin{eqnarray}
\co_1^-  \equiv  - i \sqrt{\nu} e^{i\theta_a} A \xi_{\nu-1}  e^{ i\frac{5\pi\nu}{4}}e^{-i\zeta} ,\qquad
\co_2^- \equiv   \( A e^{i\frac{3\pi\nu}{4}}   +  B e^{-i\frac{\pi\nu}{4}}\)   e^{i\zeta} ,
\label{a-}\\
\co_1^+ \equiv    i \sqrt{\nu} e^{i\theta_a} B \xi_{\nu-1}  e^{i\frac{5\pi\nu}{4}} e^{-i\zeta} ,\qquad
\co_2^+ \equiv    \(A e^{-i\frac{\pi\nu}{4}}   +  B e^{i\frac{3\pi\nu}{4}}\)   e^{i\zeta} ,
\label{a+}
\end{eqnarray}
and we used that
\be
	\frac{\Bc_{12}^{(0)}}{ \s N_1^{(0)}} \, \frac{|N_1^{(0)}|^{1/2}}{|N_2^{(0)}|^{1/2}} =\sqrt\nu e^{i \theta_a},
\ee
which follows from  \Ref{sigma}, the definition of $\theta_a$ \Ref{theta-res}, and  the choice of the branch of the square root as
\be\label{nu_gen_s}
\sqrt{\nu} = e^{i \frac{\pi}{4} {\rm sgn}(N_1 N_2 )}\sqrt{|\nu|}.
\ee.
 Due to the condition \Ref{phase-condition} we actually have $\theta_a=0$.

The formulas \Ref{a-}, \Ref{a+} enable us to find the  transition matrix $\T$, as  a solution of the equation \Ref{trMa} valid for all values of $A$ and $B$.
It reads
\be
\T
= \(\begin{array}{cc}
		 e^{ i\pi\nu} &
			\frac{i \sqrt{2\pi\nu} e^{ i\frac{\pi\nu}{2}} }{e^{2i\zeta
			} \Gamma(1-\nu)} \\
		\frac{ \sqrt{2\pi\nu}\, e^{ i\frac{\pi\nu}{2}}e^{2i\zeta
			}}{\Gamma(1+\nu)}&
			 e^{ i\pi\nu}
		\end{array}\),
\label{T}
\ee
here we used the identities
$\xi_{\nu-1}
=\frac{\sqrt{2\pi}}{\Gamma(1-\nu)} e^{-i\pi\nu}$ and
$\Gamma(1-\nu)\sin \pi \nu 
=\frac{\nu \pi}{ \Gamma(1+\nu) }$.

As it must, the obtained transition matrix  \Ref{T} satisfies general properties discussed in Appendix \ref{app:T_prop} for any phase factor $e^{i\zeta}$. To check this fact, one should take into account that $\overline{\Gamma(1+\nu)} = \Gamma(1-\nu)$ and  \Ref{nu_gen_s}.

\subsubsection{Asymptotics of $\T$ and its canonical form}\label{sec:can-matr}
First, we  check the  limit of vanishing perturbation, $\nu\to0$. We have
\be
\T
\seq_{\nu\to0} \(\begin{array}{cc}
		 1\ &
			\ 0\\
		0\ &
			\ 1
		\end{array}\)
\label{T-unit}
\ee
for any choice of $\zeta$ and both for the avoided crossing case and real turning points.

To consider the $|\nu| \gg 1$ asymptotics, we start with the Stirling's formula \cite{abramowitz}
\be
\Gamma(1+\nu) \seq \sqrt{2 \pi \nu} \( \frac{\nu}{e} \)^{\nu}, \quad  |\nu| \gg 1, \quad \arg{(\nu)} < \pi.
\ee
From the definition of $\nu$, see \Ref{nu_gen-0}, it follows that $\nu = i |\nu| \sgn (N_1   N_2)$, and we obtain
\be\label{Gam-as}
 \Gamma(1+\nu) \seq_{|\nu| \to \infty} \sqrt{2 \pi|\nu|}\,
		 e^{-|\nu|\frac{\pi}{2}}
\, e^{i  \theta_{\Gamma}(\nu)}   , \quad
	i \theta_{\Gamma}(\nu) =  - \nu + \nu \ln|\nu| +i \frac{\pi}{4}{\rm sgn}({{  N_1   N_2}} ).
\ee
Note that  $\arg{\Gamma}(1-\nu) = - \arg{\Gamma}(1+\nu)$.  Now we may simplify the off-diagonal terms  $t_{12}$ and $t_{21}$  of $\T$ as follows:
\be
t_{12} \seq
-\frac{e^{-i\frac{\pi}{2}(1-{\rm sgn} (N_1 N_2)) (1+i|\nu|)  }}
	{ e^{ \nu - \nu \ln|\nu| +2i\zeta
	}},  \qquad
t_{21} \seq
 \frac{ e^{- |\nu|\frac{\pi}{2}( {\rm sgn} (N_1 N_2)  - 1 )  }  }
	{e^{ - \nu + \nu \ln|\nu|   -2 i\zeta
	    }}.
\ee
Recalling the choice of the phase factor ${i\zeta}$ \Ref{Theta1} as
$$
	i\zeta=
	  -\frac{\nu}{2} + \frac{\nu \ln|\nu|}{2},
$$
we confirm that for the canonical modes \Ref{hWKB1-canon} the transition matrix asymptotics for $|\nu|\gg 1$ has the simplest possible form indeed
\be
\T \seq_{|\nu| \to \infty} \(\begin{array}{cc}
		 0\ & - 1 \\
		1\ & \phantom{-} 0				\end{array}\)  \quad{\rm if\ }    N_1 N_2 >0,  \qquad
\T \seq_{|\nu| \to \infty} e^{\pi |\nu|}\(\begin{array}{cc}
		   1\  & 1 \\
		 1\ &    1	
		\end{array}\) \quad {\rm if\ }     N_1   N_2 <0.
\label{T-lim}
\ee
We interpret the first of the last formulas as follows. If $\nu \gg 1$ and $N_1   N_2 >0$, then the gap between eigenvalues near degeneracy points is large enough that adiabatic approximation  found for the whole operator $\hK$  is valid. However, the matrix $\T$ is not the unit matrix. Its form
reflects the peculiarities of the numbering method and the choice of signs of adiabatic modes \Ref{hWKB1-canon} for $x \le 0$ that we adopted.  Note that the  matrix entry $-1$  can be changed at will by choosing the phase of adiabatic modes, but it would also change \Ref{T-unit}.

Substituting \Ref{Theta1} into \Ref{T} we get the final result
\be
\T = \(\begin{array}{cc}
		 e^{ i\pi\nu} &
			\frac{i \sqrt{2\pi\nu} e^{ i\frac{\pi\nu}{2} +\nu - \nu \ln|\nu| } }{\Gamma(1-\nu)} \\
		\frac{ \sqrt{2\pi\nu}\, e^{ i\frac{\pi\nu}{2} -\nu + \nu \ln|\nu|}}{\Gamma(1+\nu)}&
			 e^{ i\pi\nu}
		\end{array}\).
\label{final-T}
\ee
Note that $\sqrt{\nu} = e^{i \frac{\pi}{4} {\rm sgn}(N_1 N_2 )}\sqrt{|\nu|}$.

\vskip3mm
We present the matrix $\T$ in yet another form, writing explicitly the absolute values and arguments of matrix elements. To this end, we take into account the fact  that $|\Gamma(1+\nu)|^2 = |\Gamma(1-\nu)|^2 = \pi |\nu|/\sinh{\pi |\nu|}$ and obtain
\be
\T = \(\begin{array}{cc}
		 e^{ -\pi|\nu|w} &
			 \sqrt{1 - e^{-2 \pi |\nu|}}  e^{ -\frac{\pi}{2} |\nu|(w -1)}  e^{ i\theta'}
e^{i \frac{\pi}{2}(1+w)}  \\
		 \sqrt{1 - e^{-2 \pi |\nu|}}  e^{ -\frac{\pi}{2} |\nu|(w -1)}  e^{ -i\theta'}&
			 e^{ -\pi|\nu|w}
		\end{array}\),
\label{final-sq-T}
\ee
where
$ w=\sgn N_1 N_2 $, and
\be\label{theta-1}
\theta' = \arg{\Gamma(1+\nu)} - \theta_{\Gamma},
\ee
and $\theta_{\Gamma}$ is the principal term of the asymptotics of $\arg{\Gamma(1+\nu)}$  for $|\nu| \to \infty$, see \Ref{Gam-as}, therefore $\theta' \to 0$ in this limit.

The transition matrix $\T$ \Ref{final-T}, or equivalently \Ref{final-sq-T}, between the canonical modes \Ref{hWKB1-canon} we call \emph{canonical transition matrix}.




\section{Conclusions and physical interpretation}\label{sec:concl}

We developed an asymptotical method and studied in details the transformation of the adiabatic modes for the Schroedinger type equation \Ref{mainEq0} near a pair of degeneracy points of eigenvalues of the corresponding spectral problem. The degeneracy points appeared after small perturbation of an operator  with just one isolated point of eigenvalues' degeneracy. The main peculiarity of our statement of the problem is consideration of a linear operator pencil, which is a direct generalization of Eq. \Ref{Sch0} to non-Hermitian Hamiltonians of a special type.

The striking distinction of \Ref{mainEq0} as compared to Schroedinger equation with the self-adjoint Hamiltonian  \Ref{mainEq00}, is the absence of the positive definiteness of the corresponding conservation law, which reads $(\Ps,\G \Ps) =const$. Indeed its sign, $\sgn(\Ps,\G \Ps)=\sgn(\vph,\G\vph)$, where $\vph$ is an eigenfunction of the spectral problem, can take both positive and negative values. For problems of wave propagation known to the authors this conserved quantity has the meaning of averaged flux of energy, and therefore we called it here the flux.

The absence of the positive definiteness opens a possibility for description of two situations with the same equation -- the avoided crossing of perturbed eigenvalues, and the unavoidable crossing as well.  We studied both cases on the same footing by the method of matched asymptotic expansion with a simplifying assumption. This assumption concerns small parameters of the problem: the adiabatic parameter $\h$ and the parameter characterizing an order of the perturbation of the operator $\K$, which is assumed to be $\sh$, see \Ref{mainEq0}.  We found both the adiabatic expansion (an outer expansion) and the asymptotic expansion near the degeneracy point \Ref{conn-ps}, \Ref{conn-ph0}, where adiabatic approach is not applicable (an inner expansion). The transition matrix connecting adiabatic modes  \Ref{hWKB1-canon} to the left and to the right of degeneracy points, was presented as well in \Ref{final-T}, \Ref{final-sq-T}. The obtained matrix depends on the constant factors of adiabatic modes, which are fixed  by \Ref{phase-condition}. The assumption on small parameters helps us to avoid difficulties caused by exponentially small entries of the transition matrix.

The transition matrix depends only on eigenvalues and eigenfunctions of two degenerating modes  and does not depend on the rest of the spectrum, as is prescribed by the adiabatic theorem.   The absolute values of the canonical matrix entries, which are responsible for energy transmission, are completely determined by perturbed eigenvalues behavior. The  results obtained coincide with those of Landau and Zener \cite{landau1965collected}, \cite{zener_32} and \cite{hagedorn1991proof} for the avoided crossing scenario and \cite{reijnders2013semiclassical}, \cite{zalipPRB.91} for the two real turning points scenario.   The only restriction on the choice of modes to obtain the absolute values of the matrix entries is that they  should have the same normalization.

If, in addition, the eigenfunctions of the unperturbed operator  are known at the degeneracy point, the phases of adiabatic modes can be fixed by \Ref{phase-condition}  and the phases of the transition matrix entries  can be obtained. They coincide with those found in \cite{bobashov,bobashev1976coherent} for the case of an avoided crossing and with \cite{zalipPRB.91} for two real turning points.

\vskip 5mm
The physical interpretation of obtained results depend on the signs of the energy fluxes $N_j$, $j=1,2$ of modes being degenerated. If the signs are different, we deal with the process of reflection from the degeneracy points neighborhood area and transmission through this area. For negative $x$, the mode with positive $N$ is interpreted as an incident mode and the mode with negative $N$ as a reflected one. For positive $x$, the mode with positive $N$ is a transmitted mode.
Putting in the definition of the transition matrix \Ref{trMa} $\co_1^-=1$, $\co_2^-=R$, $\co_1^+=T$, $\co_2^+=0$, we obtain for the reflection $R$ and transmission $T$ coefficients the following:
\be
	R=-\frac{\Tc_{21}}{\Tc_{22}},\qquad
	T=\frac{\det \T}{\Tc_{22}}.
\label{rt}
\ee
If the modes are canonical  \Ref{hWKB1-canon}, the transition matrix
 \Ref{final-sq-T}  for $N_1 N_2=-1$  gives
\be
  R =- \sqrt{1 - e^{-2 \pi |\nu|}} e^{ -i\theta'}, \quad
  T = e^{- \pi |\nu|},
\ee
where $\theta'$ is defined in \Ref{theta-1}.

\vskip3mm

If the signs of energy fluxes $N_j$, $j=1,2$, are equal we have the transformation (or interaction) of adiabatic modes near the degeneracy points. For this process the chosen numbering of modes is not the most natural one. The perturbed eigenvalues approximating an original one with a given number become  nonsmooth functions, see \Ref{hb-b-t+}, \Ref{hb-b-t-}. If we change the numbering of modes on the left of the degeneracy point, then the perturbed eigenvalues for large $|\nu|$ will be continuous and the transition matrix for the new numeration reads
\be
\tilde\T
= \T \(\begin{array}{cc}
	 0  &  1 \\
         -1  &  0
         \end{array}\).
\ee
The first mode incident from the left, i.e., $\co_1^-=1$, $\co_2^-=0$ in \Ref{ps-infty-}, produces two modes on the right of $x=0$ with  $\co_1^-=\tilde{t}_{11}$, $\co_2^-=\tilde{t}_{21}$, where the corresponding elements of the matrix $\tilde\T$ are the transmission coefficient $\tilde{t}_{11}$ of the first mode,  and the excitation coefficient $\tilde{t}_{21}$  of the second one:
\be \nonumber
 \tilde{t}_{11} = \sqrt{1 - e^{-2 \pi |\nu|}} e^{ i\theta'}, \qquad \tilde{t}_{21} = e^{- \pi |\nu|}.
\ee

\vskip3mm

Finally, a comment on other methods is due. The usual way of treating the non-Hermitian problems consist of dealing with a non-Hermitian Hamiltonian $\H$ and constructing biorthogonal bases  $\{\vph\}$, $\{\vph^\dag\}$ of eigenfunctions of $\H$  and $\H^\dag$ correspondingly. In such approaches, the normalization of a solution $(\Ps^\dag, \Ps)$ can always be chosen positive, since there is no relation between  $\vph$ and $\vph^\dag$ presupposed, and the latter one can always be supplied with an appropriate phase factor. In this case, the nature of  perturbed degeneracy points (either avoided or real crossing of eigenvalues will be invoked) is fully defined by the (non-Hermitian) perturbation operator $\B$, and its matrix elements $(\vph^\dag, \B\vph)$.

The connection of our approach with the aforementioned one follows from the relation $\vph^\dag= \a \G\vph$, which is valid for the non-self-adjoint Hamiltonians factorized as $\G^{-1}(\K \,+\, \sh\B)$. We put $\a=1$, thus fixing the ambiguity in the  definition of the adjoint eigenfunction. The advantage of our approach is that the nature of the perturbed degeneracy is revealed by simple comparison of the signs of the normalization of the degenerating modes, $\sgn (\Ps_i, \G\Ps_i)=\sgn (\vph_i, \G\vph_i)$, $i=1,2$.

We believe that the suggested  approach enables one to get the results in a truly transparent manner. Our considerations can be generalized to the case where $\G^{-1}$ does not exist or $\G$ is not a matrix but, for example, a differential operator.
Further generalization of the approach to the multidimensional case, i.e., to the equation of the form of
\be
	\(\K(x,y)+\delta \B(x,y)\)\Ps= -i\h \left( \G_x\frac{\partial \Ps(x)}{\partial x}
+ \G_y\frac{\partial \Ps}{\partial y} \right)
	\label{main-multi}
\ee
is also possible, and is the subject of the future work. An example of treatment of
\Ref{main-multi} for $\K$ and $\G$ independent on $x$ and $y$  is given in  \cite{PerelSidorenko-crystal}.

The extraction of $\G$ can also reflect particular symmetries of a  given  physical problem, as is for the Dirac equation, see \Ref{dirac}. Another example of the applications of the equation in the form \Ref{mainEq0}
is the case of ${\cal PT}$-symmetric quantum mechanics with $\G$ playing the role of a ${\cal P}$-symmetry operator (see \cite{Bender2007}).
We shall note that some aspects of the presence of degeneracy points, also called \emph{exceptional} ones \cite{BerryPT04,heiss2012physics},  were considered, e.g.,  in \cite{Andrianov07,Reyes12}. However, only the case of  complex degeneracy points (i.e., avoided crossing) were dealt with, since the presence of real degeneracy points would signal entering the broken ${\cal P}$-symmetry region.
Considering the transition of a quantum system through such a region is yet another application of the general method presented here.

\section*{Acknowledgments}
This work was supported in part (I.V.F) by grant 2012/22426-7, S\~ao Paulo Research Foundation (FAPESP), (M.V.P.) grant RFBR 17-02-00217.


\appendix{}



\section{Auxiliary general facts \label{sec-aux}}

We give here  some general facts from \cite{kato2013perturbation} formulated for our case and prove two lemmas.

\begin{itemize}
\item
For our operator $\G^{-1}\K$, where $\K$ and $\G$ are selfadjoint, $\G$ and $\G^{-1}$ are bounded, we can construct a projection ${\P}$ to the invariant subspace $M'=\P \cal{H}$, corresponding to an isolate part of the spectrum $\Sigma'$. In our case, $\Sigma'$ consists of a  single eigenvalue or a pair of degenerating eigenvalues separated from  the other spectrum with a gap. All the Hilbert space $\cal{H}$ is split into the direct sum
\be
{\cal H} = M' \bigoplus M'',
\ee
where $M''=(\I-\P)\cal{H} $. Then every vector $\f$ is represented as ${\f} = {\P}{\f}  + (\I-{\P}) {\f}$,
${\P}{\f} \in M'$, $(\I-\P) {\f} \in M''$.  Both subspaces $M'$, $M''$ are invariant  for the operator $\G^{-1}\K$.

\item
The projection $\P$ can be found as
an integral of the resolvent of  $\G^{-1}\K$ along the contour $C$, which surrounds an isolated part of its spectrum $\Sigma'$:
\be
{\P} = \frac{1}{2 \pi i} \int_{C} (\G^{-1}\K -\b {\I})^{-1} \, d\b.
\ee
The adjoint projection ${\P}^+$  is an integral of the resolvent of the adjoint operator along the contour $C'$, which surrounds an isolated part of the spectrum of the adjoint operator:
\be
{\P}^+ = \frac{1}{2 \pi i} \int_{C'} ((\G^{-1}\K)^+ -\b {\I})^{-1} \, d\b.
\ee

\end{itemize}

We note that if $\K$ and $\G$ are selfadjoint, then the following is valid:
\begin{enumerate}
\item
\be
(\G^{-1}\K)^+ = \K \G^{-1}
\ee
\item
The spectrum of the adjoint operator $\K \G^{-1}$ coincides with the spectrum of the operator $\G^{-1}\K$ itself; therefore the contours coincide, $C'=C$. The adjoint eigenfunctions $\vph_j^+$ are determined as follows: $\vph_j^+ = \G \vph_j. $

\item
The projections satisfy the relations
\be\label{adjoint-op}
{\P}^+ = \G {\P} \G^{-1},
\ee
 which follows from the relation  $ \K\G^{-1} - \b \I = \G (\G^{-1}\K - \b {\I}) \G^{-1}.$

\end{enumerate}

\vskip 24pt

 {\it Lemma 1}

 The subspaces $M'$ and $M''$ are $\G$ - orthogonal, i.e.,
 \be
 ({\f},\G {\g}) = 0 \quad{\rm if\ }  {\f}\in M',  {\g}\in M'',
 \ee

 \vskip12pt
  which follows from the relations
  \begin{eqnarray}
 && {\f} = {\P}{\f}, \quad {\g} = ({\I}-{\P}){\f},\\
 && ({\P} \f, \G(\I-\P)\f) = (\f, \P^+ \G(\I-\P)\f),\\
 && \P^+ \G(\I-\P)=\G \P \G^{-1}\G(\I-\P)= \G \P(\I-\P)=0.
  \end{eqnarray}

  \vskip 24pt
  {\it Lemma 2}

  If ${\q} \ne 0$ and
  \be
 ({\f},\G {\q}) = 0
  \ee
 for all ${\f}\in M'$, then ${\q}\in M''.$

  \vskip 12pt

  Assume that $\q=\q'+\q''$, ${\q'}\in M',$ ${\q''}\in M''.$
Then
\be \label{f-G-q'}
({\f},\G {\q'}) = 0,
\ee
 because $({\f},\G {\q''}) = 0$ according to Lemma 1.
 A vector $\G \q'$ as  any vector from the Hilbert space can be represented as a sum $\G \q' = \f_1 + \g_1$, ${\f_1}\in M'$,  ${\g}_1\in M''.$
For $|\G \q'|^2$ we obtain  $|\G \q'|^2 = (\G\q', \f_1) + (\G\q', \g_1) = (\q',\G \f_1) + (\q',\G\g_1) = 0$ because of \Ref{f-G-q'} and Lemma 1.  We derive that $\q'=0$ and $\q=\q'' \in M''$.


\section{Properties of the $\G$-eigenvalue problem} \label{app:eig-pr}

We give here  properties of the  eigenvalue problem \Ref{newEVP}:
\be
	\K(x)\vph(x) = \b(x)\G\vph(x),
	\label{sEVPa}
\ee
which have distinctions for $\G \ne \I $ ($\I$ is the identity matrix), as compared with the case of $\G =\I$.

\begin{enumerate}
\item{
First of all, the eigenvalues may be both real and complex. The complex eigenvalues appear in pairs: $\b_n$ and its complex conjugate $\overline{\b_n}\equiv \b_{\n}$. The index $\n$ means the sequential number of the eigenvalue $\overline{\b_n}$.  The eigenfunctions corresponding to $\overline{\beta_n}$ are denoted by $\vph_{\n}$.}
\item{
 The normalization factor for a real eigenvalue $\b_n$
\be\label{norm-def}
N_n = (\vph_n, \G \vph_n)
\ee
is real, but may be positive and negative.
This causes new properties of solutions. We note that it is always positive
if $\G= \I$ and $\vph_n \ne 0$.

For complex eigenvalues $\b_m$ the self-orthogonality phenomena takes place and the  normalization factor is given by $(\vph_{\overline{m}}, \G \vph_m)$.}

 \item{
 The eigenfunctions of the numbers $m$ and $n$ satisfy the orthogonality conditions
\be\label{ort}
(\vph_n,\G \vph_m) =0 , \quad
{n} \ne  \overline{m}.
\ee
}

 \item{ If the eigenvalue $\b_n$ is not degenerate, we have
 \be
 N_n \ne 0.
 \ee
The same is true if $\b_n$ is twofold degenerate, and the algebraic multiplicity is equal to  the geometric one.

 }

\item{
Let $\b_1$ and $\b_2$ be  eigenvalues of the problem \Ref{sEVPa} separated from  the rest of the spectrum with a gap. The corresponding eigenfunctions are $\vph_j$, $j=1,2$.

If $(\vph_{j},\G\f)=0$, $j=1,2$, the equation
\be\label{cor-lm2}
	(\K-\b_1\G)\vph_{1\perp} =
		 \G \f
\ee
has an unique bounded solution $\vph_{1\perp}$ such that  $(\vph_{j},\G\vph_{1\perp})=0$, $j=1,2$.
}

\item{ Let $\b_n$ be real. Then
\be\label{db-dx-h}
 \frac{d \b_n(x)}{ d x} = \frac{\Kc'_{nn}(x)}{N_n}.
\ee

 Let $\b_n$ and $\b_j$ be real. Then
\be\label{conver-matr}
 (  \b_n(x) -   \b_j(x))    S_{jn}(x) =   \Kc'_{jn}(x),
\ee
where
\be\label{H1-matr}
  \K'(x) = \frac{d  \K(x)}{d x}, \quad   \Kc'_{jk}(x) \equiv (  \vph_j(x),  \K'(x)   \vph_k(x)), \quad j=1,2,\quad k=1,2,
\ee
and the conversion coefficients $S_{jk}(x)$ are defined in \Ref{S11}.

}

\item
Let $\b_1$ and $\b_2$ be real and $  \b_1(\varkappa) =   \b_2(\varkappa).$  Then
the matrix entries of $\K'$ satisfy the condition
\be\label{H12}
  \Kc'_{12} (\varkappa)= \Kc'_{21} (\varkappa) = 0.
\ee

\end{enumerate}

Now we proceed with a proof of the properties 1,2,3.
Taking the inner product of the equation (\ref{sEVPa}) for $\vph_n$ and the eigenfunction $\vph_n$, we derive
that $\b_n$ is real if $N_n \ne 0$, using the fact that  $(\vph_n,\K\vph_n)$ and $N_n$ are real. Taking the inner product
of the equation (\ref{sEVPa}) for $\vph_m$ and the eigenfunction $\vph_n$, we obtain
 \be\label{help-ort}
(\overline{\b_n}-\b_m) (\vph_n,\G \vph_m)=0.
 \ee
We used the fact that  $\K$ is self-adjoint and the property of the inner product
$(\b_n\vph_n,\G \vph_m)=\overline{\b_n} (\vph_n,\G \vph_m)$. The relation  (\ref{help-ort}) yields (\ref{ort}).

\vskip 24pt
The property 4 follows from  Lemma 2 of Appendix \ref{sec-aux}.
Consider a one-dimensional $M'$, which is $c \vph_n$ for any constant $c$. The subspace $M'$ has a single basis vector $\vph_n.$  If $(\vph_n,\G\vph_n)=0$, then $\vph_n$  belongs to $M''$ as well. We have a contradiction, because it belongs to $M'$.

 Without loss of generality $n=1$ and let $\b_1$ be a twofold degenerate eigenvalue. There are two linear independent eigenfunctions $\vph_1$ and $\vph_2$ in the corresponding invariant subspace $M'$ and
$(\vph_2,\G\vph_1)=0$. We intend to show that $N_1 \ne 0.$  We prove it by contradiction.
 Let $N_1=(\vph_1,\G\vph_1)=0$, then  $\vph_1$ belongs to $M''$ according to Lemma 2. This fact  contradicts  the assumption that $\vph_1 \in M'$.

\vskip 24pt
The property 5 follows from Lemma 2 as well. It is assumed that  $\f$ is $\G$ -- orthogonal to $M'$. Then $\f \in M''$ by Lemma 2. The subspace $M''$ is an invariant subspace for the operator $(\G^{-1}\K-\b_1 \I)$, which is invertible. The solution can be found in the form
$\vph_{1\perp}^{(n)} = (\G^{-1}\K-\b_1 \I)^{-1}_{\perp} \f$, where $(\G^{-1}\K-\b_1 \I)^{-1}_{\perp}$ is the restriction of the resolvent to $M''$, which is bounded.


\vskip 24pt
 To prove the property 6 we differentiate the eigenvalue equation (\ref{newEVP}) for the $ \vph_n$ with  respect to $x$ and scalar multiply it by $  \vph_n$. We get
\be\label{cond-help}
	( \vph_j(x),  \K'(x)  \vph_n(x)) = \(  \b_n(x)-\overline{ \b_j(x)}\) (  \vph_j(x), \G \frac {d \vph_n(x)}{d x})
+ \frac{ d \b_n(x)}{d x}   N_n \,\delta_{nk},
\ee
where $\delta_{nk}$ is the Kronecker symbol.
If $j=n$, we obtain \Ref{db-dx-h}.  If  $j\ne n$ and the eigenvalues are real, we have \Ref{conver-matr}.

\vskip 24pt
The property 7 follows from \Ref{conver-matr} if we take into account the fact that
the conversion coefficient  $  S_{jk}(x)$ is bounded for any $x$, and $  \b_1(\varkappa) =   \b_2(\varkappa)$.

\section{Perturbation method for the spectral problem}\label{sec:EVP}
We adapt here the perturbation method developed by Schroedinger for problems of Quantun Mechanics (see the history of the problem and references in \cite{kato2013perturbation}) to deal with the operator pencils, i.e., with eigenproblems containing $\G$ on the right-hand side.

 We construct an approximate solution to the eigenvalue problem both away from the degeneracy point and in its vicinity. These considerations enable us
to introduce physically relevant parameters  governing our results, and
to clarify the conditions, which cause eigenvalues behavior either as shown in Figs. \ref{compl_TP} or \ref{real_TP}.

\subsection{Perturbed eigenproblem away from degeneracy point}\label{sec:EVP-away}
Away from degeneracy point, we may search for the eigenvalues and eigenfunctions of 
the problem
\be\label{slVarEVP0}
	(\K+\sh \B )\vvph = \hb \G \vvph
\ee
as a formal asymptotic series in powers of $\sh$ and as functions of the  original variable $x$,
\be
	\hb(x,\h)={\hbn0}(x)+\sqrt\h{\hbn1}(x)+\ldots,\qquad
	\vvph(x,\h)=\vvphn0(x)+\sqrt\h \vvphn1(x)+\ldots.
	\label{anz0}
\ee
This procedure is very well studied for $\G=\I$, and will mostly be used as the reference in the rest of the paper, so we just indicate the main steps.

Upon inserting these series into \Ref{slVarEVP0} and equating terms with equal powers of $\sh$, we get an infinite sequence of equations
\begin{eqnarray}
	(\K-\hbn0\G)\vvphn0&=&
		0,\label{0-th-B}\\
	(\K-\hbn0\G)\vvphn1&=&
		 \( \hbn1\G-\B\)\vvphn0,
		\label{1-th-B}\\
	(\K-\hbn0\G)\vvphn{2}&=&
		 \( \hbn1\G-\B\)\vvphn1 + \hbn2\G\vvphn0,
		\label{2-nd-B}\\
&\ldots.&\nonumber
\end{eqnarray}
 We  supply the expansion \Ref{anz0} with condition
\be
	({\vvph}^{(0)},\G{\vvph}_j^{(n)})=0.
	\label{add-cond}
\ee

Equation \Ref{0-th-B} is the original spectral problem \Ref{newEVP}. Thus, we choose the principal approximation as
\be\nonumber
	\hb_{j}^{(0)}=\b_j, \qquad
	{\vvph}_j^{(0)} = \vph_j,
\ee
where $j$ is the number of the corresponding solution of the eigenproblem, for the sake of definiteness we choose to construct here the first mode, $j=1$.

The solution of the above system of equations differs from the standard case
only in the definition of the unperturbed eigenvalues and eigenfunctions $\b$, $\vph$, and the presence of $\G$.
 Thus, in the first approximation we can write
\be
\vvph_1=\vph_1
	+\sqrt\h\[\frac{ \Bc_{ 21}}{( \b_1- \b_2)   N_2} \vph_2
		+ \vph_{1\perp}^{(1)}\]
	+\ldots,   \qquad \qquad \qquad \qquad \qquad \qquad \qquad \quad
	\label{vvph}
\ee
\be
\hb_1= \hb_1^{(main)} + O(\h^{3/2}), \qquad
\hb_1^{(main)} = \b_1+\sqrt\h \frac{ \Bc_{ 11}}{  N_1}
 +\h\[\frac{  \Bc_{ 21}  \Bc_{ 12}}{ ( \b_1- \b_2)   N_1  N_2 }
	+\frac{ (\vph_1,\B \vph_{1\perp}^{(1)})}{  N_1} \].
	\label{hb}
\ee
We separated explicitly the contribution of the second mode $\vph_2$ since we assume that it is the pair of $\b_1$ and $\b_2$ which degenerates at $x=0$.
 By $\vph_{1\perp}^{(1)}$ we denoted the contribution of the rest of the spectrum (if any) to the first approximation of $\vvph_1$, $\G$-orthogonal to $ \vph_1$ and $ \vph_2$.
In \Ref{vvph}, \Ref{hb}, we also introduced the following notation for the  eigenfunctions $\G$- normalization
\be
	N_i=(\vph_{i},\G\vph_i), \quad i=1,2.
	\label{N-def}
\ee
It can always be chosen constant.  The matrix elements for an operator ${\bm{\mathcal A}}$ are defined as
\be
	{\cal A}_{ij}(x)\equiv \(\vph_i(x),{\bm{\mathcal A}}(x)\vph_j(x)\),\quad j,i=1,2.
	\label{matrix-ele}
\ee
The scalar product here is that inherent from the Hilbert space in which $\K$ acts.

\subsection{Perturbed eigenproblem in the vicinity of the degeneracy point}\label{sec:EVP-close}

As  is evident from \Ref{vvph}, \Ref{hb}, the expansions are not valid whenever $\b_j$, $j=1,2$, degenerate, and they do degenerate in $x=0$ according to our assumption \Ref{mb-cross}. 

In the vicinity of a degeneracy point, we describe the behavior of the eigenvalues and eigenfunctions  in terms of a \emph{slow, or stretched, variable}
\be
	\tau=x/\sh.
	\label{slVar0}
\ee
introduced in Section \ref{sec:wkb-valid}.


We substitute   (\ref{slVar0}) in $\K$ and $\B$  and expand them in formal series as follows
\be\label{expan-H-B}
 \begin{aligned}
	\K(\sh \tau)&= \Kn0    + \sh\tau \Kn1   + \h \tau^2 \Kn2+\ldots, &\\
	 \B(\sh \tau)&= \Bn0     + \sh\tau \Bn1   + \h \tau^2 \Bn2 +\ldots,&
 \end{aligned}
\ee
where
$$
	\K^{(n)}=\left.\frac1{n!} \frac{d^n \K}{d x^n}\right| _{x=0}
	, \quad \B^{(n)}=\left.\frac1{n!} \frac{d^n \B}{d x^n}\right | _{x=0}.
$$
Inserting these expansions into (\ref{slVarEVP0}), we obtain
\be
	\(\Kn0+\sh (\tau \Kn1 +  \Bn0) + \h (\tau^2 \Kn2 +  \Bn1\tau)+\ldots \) \vvph = \hb \G \vvph.
	\label{slVarEVP}
\ee
We search for solutions of this eigenproblem in the form similar to \Ref{anz0}
\be\label{anzats-pert-eig}
	\hb(\sh\tau,\h)={\vbn0}(\tau)+\sqrt\h{\vbn1}(\tau)+\ldots,\quad
	\vvph(\sh\tau,\h)=\cvphn0(\tau)+\sqrt\h \cvphn1(\tau)+\ldots
\ee
Note that in distinction to the approximations $\hb^{(n)}$, $\vvph^{(n)}$ from the previous Subsection, the approximations $\vbn{n} $, $\cvphn{n} $  are now functions of $\t$,  not $x$. We mark these groups of approximations with a tilde.

Upon inserting these series into \Ref{slVarEVP} and equating terms with equal powers of $\sh$, we get an infinite sequence of equations
\begin{eqnarray}
	(\Kn0-\vbn0\G)\cvphn0&=&
		0,\label{0-th-tau}\\
	(\Kn0-\vbn0\G)\cvphn1&=&
		 \( \vbn1\G-\tau\Kn1-\Bn0\)\cvphn0,
		\label{1-th-tau}\\
	(\Kn0-\vbn0\G)\cvphn{2}&=&
		 \( \vbn1\G-\tau\Kn1-\Bn0\)\cvphn1 \nonumber \\
      &+&   \( \vbn2\G-\tau^2\Kn2-\tau\Bn1\)\cvphn0,
		\label{2-nd-tau}\\
&\ldots&\nonumber
\end{eqnarray}
Equation \Ref{0-th-tau} is the original spectral problem \Ref{newEVP} taken at the degeneracy point $x=0$. Thus we have
\be
	\vb_{j}^{(0)} = \b_0,
\ee
where
\be
\b_0 \equiv \b_1(0)=\b_2(0), \label{b-0-def}
\ee
and  we introduced the subscript, $j=1,2$, to distinguish the two modes in what follows.
For the eigenfunctions we have
\be
{\cvph}_j^{(0)} =  \a_{j1}^{(0)}(\tau) \vph_1(0) + \a_{j2}^{(0)}(\tau) \vph_2(0), \quad j=1,2,	\label{EF}
\ee
where $ \vph_{j}(0) \equiv \lim_{x \to 0}  \vph_{j}(x)$, $j=1,2$  are linear independent eigenfunctions of the original problem. To make the definition unique we determine them by the limiting transition.

The coefficients $\a_{jk}^{(0)},$ $j,k=1,2$ are unknown at this step. They can be found from the condition of the solvability of the  equation \Ref{1-th-tau}, which is the condition of orthogonality of the right-hand side of \Ref{1-th-tau} with $ \vph_k(0)$, $k=1,2$.
It gives for both  $j=1$ and $j=2$ the same system
\be
\begin{aligned}
	(\vb_{j}^{(1)} N_1^{(0)} - \tau \Kcn1_{11} - \Bcn0_{11})\, \a_{j1}^{(0)}
		+ (-\tau \Kcn1_{12} - \Bcn0_{12}) \, \a_{j2}^{(0)}&=0,&\\
	 (-\tau \Kcn1_{21}- \Bcn0_{21}) \, \a_{j1}^{(0)}
	+(\vb_{j}^{(1)} N_2^{(0)} - \tau \Kcn1_{22} - \Bcn0_{22}) \, \a_{j2}^{(0)}&=0;&
\end{aligned}
\label{eq-eig-pert}
\ee
here, $N_j^{(0)}=N_j(0)$, $j=1,2$. The matrix elements with a superscript $(n)$, $n=0,1,$ are all taken at the degeneracy point $x=0$,
\be
\Kcn1_{jk} = (  \vph_j(0),  \K^{(1)}(0)   \vph_k(0)), \qquad
	\Bcn0_{jk} =  \Bc_{\brr jk}(0), \quad j,k=1,2.
\ee
Note that, as we show in the Appendix \ref{app:eig-pr}, the relation  $\Kcn1_{12}=\Kcn1_{21}=0$ always holds.  The diagonal matrix elements $\Kcn1_{jj}$, $j=1,2,\ldots$, are proportional to the derivatives of the unperturbed eigenvalues $\b_j'(x)$ at $x=0$ according to \Ref{db-dx-h},
\be\label{deriv-b}
\b_j'(0) = \frac{\Kcn1_{jj}}{  N_j^{(0)}}.
\ee
The condition of solvability of the system \Ref{eq-eig-pert} with respect to $\a_{j1}^{(0)}$,  $\a_{j2}^{(0)}$  is the  nullification of the determinant. It gives a quadratic equation in  $\vb^{(1)}_j$. Under an appropriate choice of the notation, its solution can always be written as
\be\label{eig-1app}
	\vb^{(1)}_{1,2}(\t) = \hb_{av}^{(1)}(\t) \pm \sqrt {(\tau + b)^2 Q^2
			+   {p^2}{\rm sgn }(N_1^{(0)} N_2^{(0)}) }.
\ee
The square root is assumed to be positive if it is real. We do not discuss the complex case.

In \Ref{eig-1app}, we used the following notation.
Half  the difference of the derivatives of the unperturbed eigenvalues at $x=0$ is denoted as
\be
	\label{not-Q}
	Q = \frac{1}{2}	\(\frac{\Kcn1_{22}}{  N_2^{(0)}}  - \frac{\Kcn1_{11}}{  N_1^{(0)}} \), \qquad  Q>0,
\ee
and the degeneracy point displacement owing to the perturbation equal for both modes reads
\be \label{not-b}
  b = \frac{1}{2Q} \( \frac{\Bcn0_{\brr 22}}{  N_2^{(0)}} - \frac{\Bcn0_{\brr 11}}{  N_1^{(0)}}\).
\ee
The parameter $p$ characterizes the degree of separation of the eigenvalues for \\$\sgn (N_1^{(0)} N_2^{(0)})$ $ \equiv \sgn (N_1N_2) =1 $.
 As shown in Appendix \ref{app:eig-pr}, the norms of eigenfunctions $\vph_j$, $j=1,2$, cannot vanish, once eigenfunctions are smooth and linearly independent on the whole interval. Then the sign of the norm of $\vph_j$, ${\rm sgn}( N_j )$, is a constant even for $N_j=N_j(x)$, and so is the product, ${\rm sgn}(N_1^{(0)} N_2^{(0)}) = {\rm sgn}(N_1 N_2) $. In what follows we always use the later expression.
 or the width of the classically forbidden zone if $N_1 N_2 <0$.
It reads
\be
p^2 = \frac{\Bcn0_{12} \Bcn0_{21}}{|N_1^{(0)} N_2^{(0)}|}
	= \frac{|\Bcn0_{12}|^2}{|N_1^{(0)} N_2^{(0)}|}.
	\label{not-p}
\ee
Finally, the average of degenerating eigenvalues in the first-order approximation is as follows:
\be\label{b-av}
\hb_{av}^{(1)}(\t)
	= \frac{1}{2} \(\frac{\Bcn0_{11}}{  N_1^{(0)}} + \frac{\Bcn0_{\brr 22}}{  N_2^{(0)}}\)
		+ \frac{\t}{2} \( \frac{\Kcn1_{11}}{  N_1^{(0)}} + \frac{\Kcn1_{22}}{  N_2^{(0)}} \).
\ee

We may give now formulas for  two first principal approximations of eigenvalues $\vb_j^{(main)}$, $j=1,2$:
\be\label{eig-root}
\begin{aligned}
&\vb_{j}(\sh\t,\h) =  \vb_j^{(main)}(\sh\t,\h) + O( \h), \quad \\ &\vb_j^{(main)} = \b_0 + \sh \( \,\vb_{av}^{(1)}\, + \,
   (-1)^{j+1} \,  \sqrt {(\tau + b)^2 Q^2
		+{p^2}\,\sgn (N_1N_2)  }\, \),
 \end{aligned}
 \ee
 which are important for leading terms of adiabatics.
At this stage we are ready to deduce that for $\sgn(N_1 N_2 )=1$ we have the avoided crossing case, see Fig. \ref{compl_TP}, with two complex degeneracy points, $\varkappa_\pm$,  while for $\sgn (N_1 N_2 ) =-1$  we have two real ones, as in Fig. \ref{real_TP},
\be
\varkappa_{\pm}
	=\left\{
	\begin{array}{ll}
		\sh(-b \pm i p/Q),\ &N_1 N_2 >0\\
		\sh(-b \pm p/Q), &N_1 N_2 <0
		\end{array}
		\right.
		\label{deg-point}
\ee
In both cases the degeneracy points are the simple ones.
In what follows we use notation
\be
\t_{\pm} = \Re {\varkappa_{\pm}}\,/ \, \sh.
\label{t_pm}
\ee


The eigenfunction approximation ${\cvph}_j^{(0)}$ \Ref{EF} for $j=1,2$ can be easily found now by solving \Ref{eq-eig-pert} for $\a_j^{(0)}$, $j=1,2$, which can be written in the notation \Ref{not-Q}, \Ref{not-b}, \Ref{b-av}  as
\be\label{al-s}
	 \a_{j1}^{(0)}(\t) = \frac{\Bcn0_{12}}{N_1^{(0)} },\qquad
		 \a_{j2}^{(0)}(\t) = \vb_{j}^{(1)}-\hb_{av} + Q(\tau + b).
\ee

\subsection{Matching of eigenvalues expansion away and near the degeneracy point}
An asymptotics of $\vb_{j}^{(main)}(\sh\t,\h)$, see  \Ref{eig-root},  as $|\t| \to \infty$ reads
\be
\label{lar-t-as}
 \vb_j^{(main)}(\sh\tau,\h) \seq
\b_0
	+ \sh \(\hb_{av}^{(1)}(\t)
		+ (-1)^{j+1}  \( Q\,|\t + b| -\frac{i \nu}{|\t| } \) + o(\t^{-1})  \).
\ee

We neglect $b$ in comparison with $\t$ in the denominator of the second term. The parameter $\nu$, see \Ref{nu_gen-0} is a dimensionless combination of the above mentioned physical parameters, which will govern our final result.
The branch of $\sqrt{\nu}$ is fixed in
\Ref{nu_gen_s}.

The obtained asymptotics matches the asymptotics of eigenvalues away from the degeneracy point $\hb_j^{(main)}(\sh \t)$ if $\tau<0$, $|\tau| = \h^{-\gamma}$, $0<\gamma<1/2$.  Indeed, rewriting $\hb_j(x)$, see \Ref{hb}, in terms of the parameter $\tau$ and inserting there \Ref{mb-cross} in the form
 \be
\b_2(\sh \,\t) -  \b_1(\sh \,\t) =   2Q \sh \tau + O(\h \tau^2), \qquad  \label{del-b}
\ee
we obtain for the first mode
\be
\hb^{(main)}_1(\sh\t,\h) = \b_1(0) + \b_1'(0) \sh\t + \sh \frac{ \Bcn0_{11}}{  N_1^{(0)}} - \sh \frac {\Bcn0_{21}  \Bcn0_{12} } {2 Q \t}
+ O(\h^{1-2\gamma}).
\ee
Taking into account the notation \Ref{b-0-def}, \Ref{deriv-b}, \Ref{not-Q},
  \Ref{not-b}, \Ref{nu_gen-0}, \Ref{b-av} and also \Ref{del-b},    we obtain for $\tau<0$
\be
\begin{aligned}\label{hb-asym}
\hb_1^{(main)}(\sh\t,\h) =  \b_0 + \sh \( \hb_{av}^{(1)}(\t) -   Q (\t + b) +i \frac {\nu } {\t }\)  + O(\h^{1-2\gamma}).
\end{aligned}
\ee
Analogous considerations  for the second mode yield a coincidence with \Ref{lar-t-as}.

Formulas \Ref{hb-asym} and \Ref{lar-t-as} show that the asymptotics of the principal terms of eigenvalues away and near the degeneracy point match for $x=\sh \t$, $\t = \h^{-\gamma}$, $0<\gamma<1/2$ to the left of $x=0$
\be\label{hb-b-t+}
	\hb_j^{(main)}(\sh \tau, \h) \mathop=_{\tau\to-\infty} \vb_j^{(main)}(\sh\tau,\h) + O(\h^{1-2\gamma}), \qquad \quad j=1,2.
\ee
While to the right, the numbering of eigenvalues $\vb_j$ and $\hb_j$ is inverted
\be\label{hb-b-t-}
\begin{aligned}
	&\hb_1^{(main)}(\sh \tau, \h) \mathop=_{\tau \to +\infty} \vb_2^{(main)}(\sh\tau,\h) + O(\h^{1-2\gamma}), \qquad \\
    &\hb_2^{(main)}(\sh \tau, \h) \mathop=_{\tau \to +\infty} \vb_1^{(main)}(\sh\tau,\h) + O(\h^{1-2\gamma}).
\end{aligned}
\ee
We use these formulas as a definition of $\hb_j^{(main)}(\sh \tau, \h), j=1,2$  near  the degeneracy points.
  Note that for $N_1 N_2 <0$,   these functions are not defined on the interval $(\sh\,\t_-,\sh\,\t_+)$. In the opposite case, $N_1 N_2 >0$, we obtain that $x=\sh \,\tau_{\pm} =-\sh b$, $\t_\pm$ given by \Ref{t_pm} and $\hb_j^{(main)}(\sh \tau, \h), j=1,2$ are defined around such $x$ and have there discontinuities.


\section{ The properties of the transition matrix}\label{app:T_prop}
The general properties of the transition matrix follow from the flux conservation law \Ref{conser}.
%
For two solutions $\Ps_j$, $j=1,2$ having $\Ps_{j-}^{(0)}$ as their asymptotics to the left of the degeneracy point
\be
	\Ps_j \seq_{ {x \ll -\sh}}  \Ps_{j-}^{(0)} +O(\sh),
	\label{PS-}
\ee
their asymptotics to its right can be derived via \Ref{trMa} and will be given by
\be
	\Ps_j \seq_{ {x \gg \sh}} t_{j1}  \Ps_{1+}^{(0)}+t_{j2} \Ps_{2+}^{(0)}+O(\sh),
	\label{PS+}
\ee
where $t_{jk}$, $j,k=1,2,$ are the $\T$ matrix entries. At the same time, the relation
\be
	(\Ps_j, \G \Ps_k) = const , \quad
	j,k=1,2
	\label{const}
\ee
must hold and the constants must be the same on both sides of the degeneracy point.

Calculating  the scalar product in \Ref{const} for all values of the indices by using \Ref{PS-} and \Ref{PS+} for both sides of the degeneracy point, and equating the results, we obtain
\begin{eqnarray}
   N_1  = |t_{11}|^2   N_1 + |t_{21}|^2   N_2, \label{T-1}\\
   N_2  = |t_{12}|^2   N_1 + |t_{22}|^2   N_2, \label{T-2}\\
0 = \overline{t_{11}} t_{12}   N_1 + \overline{t_{21}} t_{22}   N_2.  \label{T-3}
\end{eqnarray}
Equation \Ref{T-3} yields
\be\label{not-kappa}
\frac{\overline{t_{11}}}{t_{22}} = - \frac{  N_2}{   N_1} \frac{\overline{t_{21}}}{t_{12}} \equiv \gamma,
\ee
where we introduce the notation $\gamma$. The last formula enables us to derive from \Ref{T-1} and \Ref{T-2} that
$\det{\T} = {1}/{ \gamma} = \overline{\gamma}.$  Therefore $|\gamma|=1$, i.e., $\gamma$ is a phase factor.
We can govern $\gamma$ by changing lower limits of integration in adiabatic formulas and may  get $\gamma=1$.
Then we deduce that for an appropriate choice of the arbitrary phase of the adiabatic mode, the matrix $\T$ has the following properties:
\be\label{gen-prop-T}
{\det{\T}} =1, \quad
\T = \(\begin{array}{cc}
		 t_{11} &  t_{12}\phantom{\Big|} \\
	- \overline{t_{12}} \frac{  N_1}{  N_2}\
		& \overline{t_{11}}\phantom{\Big|}
		\end{array}\).
\ee
If ${  N_1}={  N_2}$, the matrix $\T$ is unitary $\T \T^+= \I,$ where $\I$ is the identity matrix.


\section{Estimates for approximations of inner expansion} \label{app:est-app-inn}

Our aim here is to give a sketch of the proof of estimates
\be\label{hi-ord-estim}
a_1^{(n)}=O(\tau^{3n}),\qquad a_2^{(n)}=O(\tau^{3n})
\ee
for large $\tau$.
To obtain the estimates (\ref{hi-ord-estim}) we derive the representation
\begin{eqnarray}
a_1^{(n)} &=&  {\cal P}^+_{3n}(t)\, D_{\nu-1}(t) + {\cal Q}^+_{3n}(t)\, D_{\nu}(t) + {\cal P}^-_{3n}(t)  D_{\nu-1} (-t) +  {\cal Q}^-_{3n}(t)\, D_{\nu}(-t), \label{a1-n-pol}\\
a_2^{(n)} &=&  {\cal M}^+_{3n}(t)\, D_{\nu}(t) + {\cal N}^+_{3n-1}(t)\, D_{\nu-1}(t) + {\cal M}^-_{3n}(t)  D_{\nu} (-t) +  {\cal N}^-_{3n-1}(t)\, D_{\nu-1}(-t), \label{a2-n-pol}
\end{eqnarray}
where ${\cal P}^+_{\alpha}(t),$ ${\cal Q}^+_{\alpha}(t),$ ${\cal P}^-_{\alpha}(t),$ ${\cal Q}^-_{\alpha}(t)$ denote polynomials of degree $\alpha$, $D_{\nu}(t)$ is the parabolic cylinder function.

  The coefficients  $a_j^{(n)}$, $j=1,2$ satisfy a system
\be\label{n-app-eq}
\begin{aligned}
	& 	-i \dot a_1^{(n)} +
		(\tau +b)Q   a_1 ^{(n)}		-   \frac{B_{12}^{(0)}}{N_1^{(0)}}   a_2^{(n)}  = f_1^{(n)},
	\\
	&	-i \dot a_2 ^{(n)}-
		\frac{B_{21}^{(0)}}{N_2^{(0)}} a_1 ^{(n)}		- (\tau +b) Q  a_2^{(n)} =f_2^{(n)}.
\end{aligned}
\ee
For $n>1$ the main terms on the right-hand sides of the equations are
\begin{eqnarray}
&&f_1^{(n)} \sim \t^2 {\cal H}^{(2)}_{11}a_1^{(n-1)} +  \t^2 {\cal H}^{(2)}_{12}a_2^{(n-1)} + \ldots, \label{f1-def}\\
&&f_2^{(n)} \sim \t^2 {\cal H}^{(2)}_{21}a_1^{(n-1)} +  \t^2 {\cal H}^{(2)}_{22}a_2^{(n-1)} + \ldots. \label{f2-def}
\end{eqnarray}
Omitted terms contain $a_j^{(k)}$, $j=1,2$, $k < n-1$.
If $n=1$,  formulas (\ref{f1-def}), (\ref{f2-def}) are exact and no terms are omitted.
The equations above follow from the solvability conditions for equations of each approximation of the inner expansion.

To solve  system (\ref{n-app-eq}), we  express $a_1^{(n-1)}$ from the second equation in the form of
\be
	a_1^{(n)}= -\frac{N_2^{(0)}}{B_{21}^{(0)}}\, \(i \dot a_2^{(n)} + (\tau +b) Q a_2^{(n)}
+ i \dot{f}_2^{(n)} - Q(\t + b) {f}_2^{(n)}\) 
	\label{a1-n}
\ee
and substitute it into the first equation of (\ref{n-app-eq}). We obtain
\be
	 \ddot a_2^{(n)}  + \((\tau +b)^2 Q^2 - i Q + \nu \s^2 \)a_2^{(n)}  =
- \frac{B_{21}^{(0)}}{N_2^{(0)}}f_1^{(n)} + i \dot{f}_2^{(n)} - Q(\t + b) {f}_2^{(n)}.
	\label{a2-n}
\ee

The formulas (\ref{a1-n-pol}), (\ref{a2-n-pol}) can be proved by the method of  mathematical induction by applying  the following fact:
the operators
$-i \frac{d}{d \t} +	(\tau +b)Q$ and $-i \frac{d}{d \t} -	(\tau +b)Q$
act on $D_{\nu}(t)$, $t=\sigma (\tau +b)$ as
\begin{equation}
\left( i \frac{d}{d \t}  +	(\tau +b)Q \right) D_{\nu}(t) = i \nu \sigma D_{\nu-1}(t), \quad
\left( -i \frac{d}{d \t} +	(\tau +b)Q \right) D_{\nu-1}(t) = i \sigma D_{\nu}(t).
\label{2-help}
\end{equation}
Generally speaking, $f_2^{(n)}$ contains the same degree of $\t$ in polynomials multiplied by $D_{\nu}(t)$ and by $D_{\nu-1}(t)$: $f_1^{(n)} = O(\t^{3n-1})$, $f_2^{(n)} = O(\t^{3n-1})$.
 From  property (\ref{2-help}) we derive that the operator on the right-hand side of (\ref{2-help}) acting on $f_2^{(n)}$ reduces by one the degree of the polynomial multiplied by $D_{\nu-1}(t)$   and increases by one the degree of the polynomial multiplied by $D_{\nu}(t)$. Because of the term $f_1^{(n)}$ the power of polynomial multiplied by $D_{\nu-1}(t)$ on the right-hand side of (\ref{a2-n}) is not changed.

 We have a natural representation  $a_2^{(n)}(\t) = a(\t) + a^-(\t)$, where $a(\t)$ satisfies
 \be\label{aa+}
	 \ddot a  + \left((\tau +b)^2 Q^2 - i Q + \nu \s^2 \right)a  =
{\cal P}_{3n-1}(\t)\, D_{\nu-1}(t) + {\cal Q}_{3n}(\t)\, D_{\nu}(t),
\ee
and the equation for $a^-(\t)$ differs from (\ref{a+}) by replacing $t$ by $-t$.

The factorisation
\be\nonumber
 \frac{d^2}{d \t^2} + \((\tau +b)^2 Q^2 - i Q  \) =  \(-i \frac{d}{d \t} +	(\tau +b)Q\) \(i \frac{d}{d \t} +	(\tau +b)Q\)
\ee
and the properties of operator factors enable us to show that a solution of (\ref{aa+}) has the form $a = {\cal N}_{3n-1}^+(\t)\, D_{\nu-1}(t) + {\cal M}_{3n}^+(\t)\, D_{\nu}(t)$. Analogous considerations for $a^-$ yield (\ref{a2-n-pol}).
Properties of (\ref{2-help}) and (\ref{a1-n}) result in (\ref{a1-n-pol}).


\bibliography{citations_tmp}{}
\bibliographystyle{unsrt}

\end{document}